\documentclass[12pt]{article}
\usepackage{epsfig, amsmath, amssymb}
\usepackage{graphicx,psfrag}
\usepackage{xcolor}
\newcommand{\nn}{\nonumber}
\newcommand{\be}{\begin{equation}}
\newcommand{\ee}{\end{equation}}
\newcommand{\ben}{\begin{equation}}
\newcommand{\een}{\end{equation}}
\newcommand{\bea}{\begin{eqnarray}}
\newcommand{\eea}{\end{eqnarray}}
\newcommand{\bA}{\begin{array}}
\newcommand{\eA}{\end{array}}
\newcommand{\bc}{\begin{center}}
\newcommand{\ec}{\end{center}}
\newcommand{\al}{\alpha}

\newcommand{\ra}{\rightarrow}
\newcommand{\del}{\partial}

\newcommand{\ie}{{\it i.e.}}
\newcommand{\eg}{{\it e.g.}}

\def\BZ{{\mathbb Z}}

\newcommand{\lan}{\langle}
\newcommand{\ran}{\rangle}

\begin{document}


\begin{titlepage}

%

\bc

\hfill 
\\         [25mm]

{\Huge Further remarks on de Sitter space,
 \\ [2mm]
  extremal surfaces and time entanglement} 
\vspace{16mm}

{\large K.~Narayan} \\
\vspace{3mm}
{\small \it Chennai Mathematical Institute, \\}
{\small \it H1 SIPCOT IT Park, Siruseri 603103, India.\\}

\ec
\vspace{35mm}

\begin{abstract}
  We develop further the investigations in arXiv:2210.12963 [hep-th]
  on de Sitter space, extremal surfaces and time entanglement.  We
  discuss the no-boundary de Sitter extremal surface areas as certain
  analytic continuations from $AdS$ while also amounting to space-time
  rotations. The structure of the extremal surfaces suggests a
  geometric picture of the time-entanglement or pseudo-entanglement
  wedge. We also study some entropy relations for multiple subregions.
  The analytic continuation suggests a heuristic
  Lewkowycz-Maldacena formulation of the extremal surface areas.  In
  the bulk, this is now a replica formulation on the Wavefunction
  which suggests interpretation as pseudo-entropy. Finally we also
  discuss aspects of future-past entangled states and time
  evolution.
\end{abstract}


\end{titlepage}

{\tiny 
\begin{tableofcontents}    
\end{tableofcontents}
}


\vspace{-5mm}

\section{Introduction}\label{sec:Intro}

Understanding holography
\cite{Maldacena:1997re,Gubser:1998bc,Witten:1998qj} in de Sitter space
(and cosmology more generally) is of great interest. Taking the far
future (or past) as the natural asymptotics leads to $dS/CFT$ with a
hypothetical non-unitary dual Euclidean CFT at the future boundary
$I^+$ \cite{Strominger:2001pn,Witten:2001kn,Maldacena:2002vr} (and
\cite{Anninos:2011ui} for higher spins). See \eg\
\cite{Spradlin:2001pw}, \cite{Anninos:2012qw}, \cite{Galante:2023uyf},
for some reviews of various aspects here.

Certain attempts at realizing de Sitter entropy \cite{Gibbons:1977mu}
as some sort of holographic entanglement entropy lead to
generalizations of the Ryu-Takayanagi formulation of holographic
entanglement in $AdS$
\cite{Ryu:2006bv,Ryu:2006ef,HRT,Rangamani:2016dms}, pertaining to
RT/HRT extremal surfaces anchored at the future boundary $I^+$ in de
Sitter space
\cite{Narayan:2015vda,Narayan:2015oka,Sato:2015tta,Miyaji:2015yva,
  Narayan:2017xca,Narayan:2020nsc,Doi:2022iyj,Narayan:2022afv}.  These
generalizations amount to considering the bulk analog of setting up
entanglement entropy in the dual CFT at the future boundary. Analysing
the extremization reveals that surfaces anchored at $I^+$ do not
return to $I^+$.  In entirely Lorentzian de Sitter spacetime, this
leads to future-past timelike surfaces stretching between $I^\pm$:
apart from an overall $-i$ factor (relative to spacelike surfaces in
$AdS$) their areas are real and positive. With a no-boundary type
boundary condition, the top half of these timelike surfaces joins with
a spacelike part on the hemisphere giving a complex-valued area
\cite{Doi:2022iyj}, \cite{Narayan:2022afv}\ (see
\cite{Hikida:2022ltr,Hikida:2021ese} for $dS_3/CFT_2$). The real part
of the area arises from the hemisphere and is precisely half de Sitter
entropy.  Some of these structures
\cite{Narayan:2017xca,Narayan:2020nsc} are akin to space-time
rotations from $AdS$ (in particular \cite{Hartman:2013qma}) and, in
analogy with \cite{Maldacena:2001kr}, suggest dual future-past
thermofield-double type entangled states (see also
\cite{Arias:2019pzy}, \cite{Arias:2019zug}, as well as
\cite{Cotler:2023xku}, \cite{Cotler:2022weg}).

More recently there has been considerable interest towards
understanding the areas of these extremal surfaces as encoding
pseudo-entropy or ``time-entanglement'' \cite{Doi:2022iyj},
\cite{Narayan:2022afv}, entanglement-like structures involving
timelike separations. Pseudo-entropy \cite{Nakata:2020luh} is the
entropy based on the transition matrix $|f\ran\lan i|$
regarded as a generalized density operator (for further developments,
including aspects of pseudo-entropy in non-gravitational theories, see
\eg\ \cite{Mollabashi:2020yie}-\cite{Omidi:2023env}). In some sense
this is perhaps the natural object here since the absence of $I^+\ra I^+$
returns for extremal surfaces suggests that extra data is required in
the interior, somewhat reminiscent of scattering amplitudes
(equivalently the time evolution operator), and of
\cite{Witten:2001kn} viewing de Sitter space as a collection of
past-future amplitudes. This is also suggested by the $dS/CFT$
dictionary $Z_{CFT}=\Psi_{dS}$\ \cite{Maldacena:2002vr}: boundary
entanglement entropy formulated via $Z_{CFT}$ translates to a bulk
object formulated via the Wavefunction $\Psi_{dS}$ (a single ket,
rather than a density matrix), leading (not surprisingly) to
non-hermitian structures.

In this paper, we develop further various aspects of the discussions
in \cite{Narayan:2022afv}. We obtain an understanding of the
future-past and no-boundary surfaces in terms of analytic
continuations from RT/HRT surfaces in $AdS$ (in part overlapping with
some of the discussions in \cite{Doi:2022iyj}): the key point is that
these analytic continuations also amount to space-time rotations (in
some sense akin to turning the $AdS$ Penrose diagram sideways). We
study this in detail first for the IR surface for maximal subregions
at $I^+$\ (sec.~\ref{sec:dSacIR}): this then paves the way to
nonmaximal subregions as well (sec.~\ref{sec:AdS3dS3}). The analysis
is straightforward to carry out explicitly for $dS_3$ (overlapping in
part with \cite{Hikida:2022ltr}) but it is also possible to identify
similar features for higher dimensional $dS_{d+1}$ although solving
exactly is difficult\ (sec.~\ref{sec:dSd+1}). The formulation via
analytic continuations suggests a natural way to obtain the
no-boundary de Sitter extremal surface areas through a heuristic
replica argument (sec.~\ref{sec:dSrepLM}) involving an analytic
continuation of the Lewkowycz-Maldacena formulation
\cite{Lewkowycz:2013nqa} in $AdS$ to derive RT entanglement entropy
(generalized in \cite{Dong:2016hjy}, \cite{Dong:2016fnf}, 
\cite{Dong:2013qoa}; see also
\cite{Casini:2011kv} and the reviews \cite{Rangamani:2016dms} and
\cite{Callebaut:2023fnf}). The crucial difference here is that since
the analytic continuation maps $Z_{bulk}^{AdS}$ to the de Sitter
Wavefunction $\Psi_{dS}$, this is now a replica formulation on the
bulk $dS$ Wavefunction. In particular the codim-2 brane that smooths
out potential bulk (orbifold) singularities is now a time-evolving,
part Euclidean, part timelike, brane. This leads to a complex area
semiclassically, with the real part arising from the Euclidean
hemisphere and the timelike part pure imaginary.

Drawing out the extremal surfaces geometrically also leads to a
version of subregion-subregion duality, just geometrically
(sec.~\ref{sec:subregDuality}). This is arrived at by defining the
bulk subregion as an appropriate ``time-entanglement'' or
``pseudo-entanglement'' wedge obtained as the appropriate domain of
dependence bounded by the boundary subregion at $I^+$ and the extremal
surface). There are various interesting differences from $AdS$ not
surprisingly, the areas being complex, so the analogs of entropy
relations for multiple subregions (mutual information, tripartite
information, strong subadditivity) show new features (sec.~\ref{sec:MITISSB}).

We also explore certain aspects of future-past entangled states and
time evolution in quantum mechanics (sec.~\ref{sec:fpTFD}), which
suggest close relations between the existence of the time evolution
operator and that of future-past entangled states.
Sec.~\ref{sec:Disc} contains some conclusions (while
sec.~\ref{rev:dSfpnb} contains a brief review of \cite{Narayan:2022afv}).

\subsection{A brief review}\label{rev:dSfpnb}

We briefly review \cite{Narayan:2022afv} here.
These generalizations of RT/HRT extremal surfaces to de Sitter
space involve considering
the bulk analog of setting up entanglement entropy in the dual
Euclidean $CFT$ on the future boundary \cite{Narayan:2015vda},
restricting to some boundary Euclidean time slice as a crutch,
defining subregions on these slices, and looking for extremal surfaces
anchored at $I^+$ dipping into the holographic (time)
direction. Analysing this extremization shows that there are no
spacelike surfaces connecting points on $I^+$: surfaces anchored at
$I^+$ do not return to $I^+$, \ie\ there is no $I^+\ra I^+$ turning
point.
 This implies that surfaces starting at $I^+$ continue inward,
to the past, thus requiring extra data or boundary conditions in the
interior, or far past. In entirely Lorentzian de Sitter space, this
leads to future-past timelike surfaces stretching between $I^\pm$
\cite{Narayan:2017xca,Narayan:2020nsc}: these are akin to rotated
analogs of the Hartman-Maldacena surfaces \cite{Hartman:2013qma} in
the eternal $AdS$ black hole. Apart from an overall $-i$ factor
(relative to spacelike surfaces in $AdS$) their areas are real and
positive. Alternatively, we could consider no-boundary de Sitter space
in accord with the Hartle-Hawking no-boundary prescription, joining
the top Lorentzian $dS$ half to a hemisphere in the bottom.  With this
no-boundary condition, the top half of the timelike future-past
surfaces glues, with regularity at the mid-slice, onto a spacelike
part that goes around the hemisphere (thus turning around), giving a
complex-valued area (see \cite{Hikida:2022ltr,Hikida:2021ese} for
$dS_3$). The top part of the surface (in the Lorentzian de Sitter) is
the same as in the entirely timelike surfaces above: this reflects
consistency of the future-past surfaces with Hartle-Hawking boundary
conditions.  The finite real part of the area of the no-boundary
surfaces arises from the hemisphere and is precisely half de Sitter
entropy. Overall, these $dS$ surfaces and their areas can be regarded
as space-time rotations from $AdS$.
The complex-valued entropies here are also reminiscent of related
objects arising from timelike-separated quantum extremal surfaces
\cite{Chen:2020tes}, \cite{Goswami:2021ksw}, as well as entanglement
entropy in ghost-like theories \cite{Narayan:2016xwq,Jatkar:2017jwz}.

\begin{figure}[h]
\hspace{2mm}
\begin{minipage}[b]{7pc}
  \includegraphics[width=6pc]{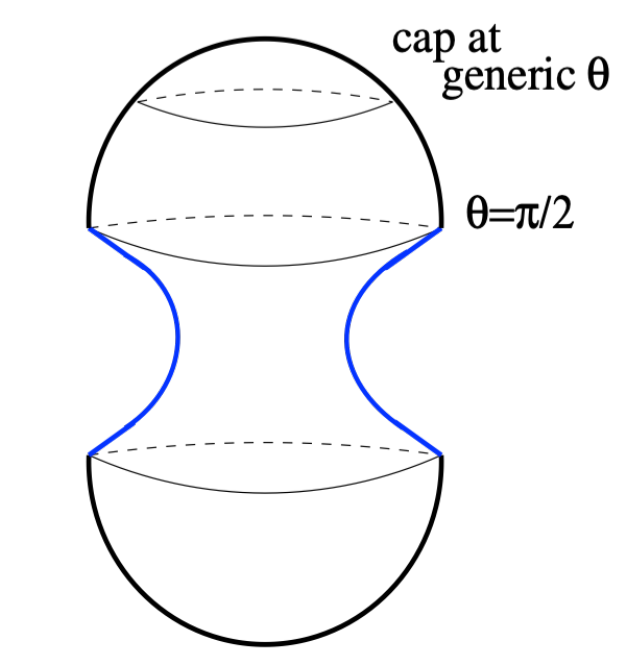}
\end{minipage}
\hspace{3mm}
\begin{minipage}[b]{22pc}
  \caption{\label{fpnb}\ {\footnotesize
      (reproduced from \cite{Narayan:2022afv})}\\
    {\footnotesize Entirely Lorentzian $dS$, future-past extremal surfaces
      (left);\\ No-boundary $dS$, no-boundary extremal surfaces (right).
      \newline}}
\end{minipage}
\begin{minipage}[b]{8pc}
\includegraphics[width=7pc]{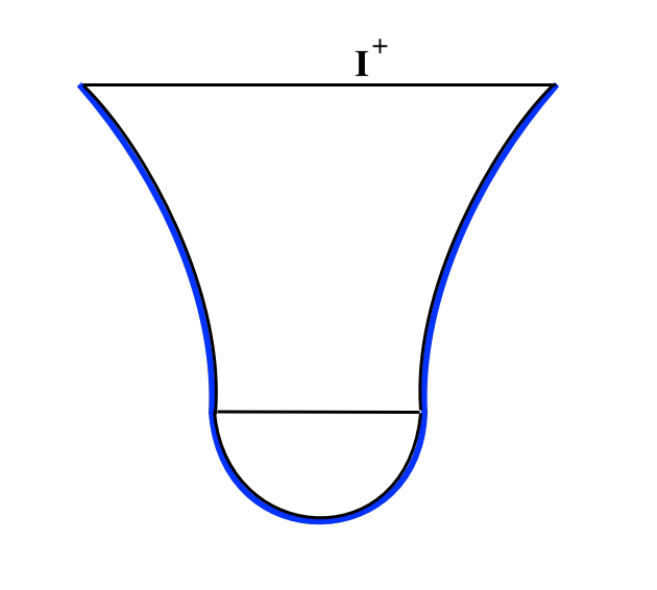}    
\end{minipage}
\end{figure}
These were further refined in \cite{Narayan:2022afv}, as well as
\cite{Doi:2022iyj}, which we develop further here. To summarize some
key points (see Figure~\ref{fpnb}):\ the areas of the future-past
and no-boundary surfaces
are of the form\ $S_{fp} = -2iS_0\,I[\tau_{_{cF}},\tau_*]$\
and\ $S_{nb} = -iS_0\,I[\tau_{_{cF}},\tau_*] + {S_0\over 2}$,\ where 
$I$ is a ``reduced'' time integral.\ 
In the IR limit when the subregion at $I^+$ is maximal, 
we obtain the extremal surface areas
\bea\label{dS43rev}
&& (dS_4)\quad S_{fp} = -i\,{\pi l^2\over 2G_4} {l\over\epsilon}\,;\qquad
S_{nb} = -i\,{\pi l^2\over 4G_4} {l\over \epsilon} + {\pi l^2\over 2G_4}
= -i\,\big( {\pi l^2\over 4G_4} {l\over\epsilon} +
i {\pi l^2\over 2G_4} \big)\quad [S_0={\pi l^2\over G_4}]\,; \nn\\ [1mm]
&& (dS_3)\quad S_{fp} = -i{l\over G_3}\log {l\over\epsilon};\qquad
S_{nb} = -i\big({c\over 3}\log {l\over\epsilon}\!
+\! {c\over 6} i\pi\big)\quad [c={3l\over 2G_3}]\,.
\eea
These $dS$ extremal surfaces at $I^+$ and their areas are analogous to
space-time rotations from $AdS$: \eg\ $dS$ future-past surfaces are
analogous to rotated Hartman-Maldacena surfaces \cite{Hartman:2013qma}
in the $AdS$ black hole. The overall $-i$ reflects the space-time
rotation while the expression inside the brackets is the
entanglement entropy for a timelike subregion in $AdS$
\cite{Doi:2022iyj}, \cite{Narayan:2022afv}.
Roughly two copies of the no-boundary surfaces glued with
appropriate time-contours make up the future-past surfaces: 
the areas of the IR extremal surfaces satisfy
\be\label{dS43rev2}
S_{fp}=S_{nb}-S_{nb}^*\,,\qquad
{\rm Re}(S_{nb}) = {1\over 2}\cdot dS\ entropy\,,
\ee
encapsulating\  $S_{nb} \equiv \Psi_{dS}$ (one copy)\ while\
$S_{fp} \equiv \Psi_{dS}^*\Psi_{dS}$\ (two copies, $I^+\cup I^-$).
$S_{nb}$ is the time-entanglement entropy, or pseudo-entropy,
in one copy $\Psi_{dS}=Z_{CFT}$: we will study the no-boundary
surfaces further in what follows.

\section{de Sitter extremal surfaces,\ analytic continuations,\
  space-time rotations}\label{sec:dSac}

The $dS$ areas (\ref{dS43rev}) were obtained in \cite{Narayan:2022afv}
and earlier work
\cite{Narayan:2015vda,Narayan:2015oka,Narayan:2017xca,Narayan:2020nsc}
by studying extremal surfaces anchored at the future boundary $I^+$ in
de Sitter directly, as reviewed above. On general grounds we expect
that at least certain quantities in de Sitter space can be understood
as appropriate analytic continuations from $AdS$, as \eg\ elucidated
long back in the reformulation \cite{Maldacena:2002vr} of the $dS/CFT$
dictionary from $AdS$. In particular boundary correlation functions
(obtained from $Z_{CFT}$) can be recast as analytic continuation,
while bulk expectation values cannot be (they involve
$|\Psi_{dS}|^2$), although there are intricate interrelations of
course.  Since the $dS$ extremal surface areas arose from bulk analogs
of setting up entanglement entropy in the boundary CFT as reviewed at
the beginning of sec.~\ref{rev:dSfpnb}, one might expect that they can
in fact be recast via some analytic continuation of extremal surfaces
in $AdS$.  In sec.~\ref{sec:dSac}, we demonstrate this, recasting
no-boundary $dS$ extremal surfaces anchored at $I^+$ via certain
analytic continuations from $AdS$, in the process illustrating how the
analytic continuations also amount to space-time rotations. A detailed
study for the IR extremal surfaces (maximal subregions) appears in
sec.~\ref{sec:dSacIR}, recovering the areas (\ref{dS43rev}) via
analytic continuation.  In sec.~\ref{sec:AdS3dS3}, we study general
subregions and $dS$ extremal surfaces in complete detail in $dS_3$,
and perturbatively in higher dimensional $dS_{d+1}$ in
sec.~\ref{sec:dSd+1}: this unifies the direct de Sitter calculation
and analytic continuation from $AdS$, tying together the discussions
in
\cite{Narayan:2015vda,Narayan:2015oka,Narayan:2017xca,Narayan:2020nsc,Narayan:2022afv}.

The analytic continuation from Euclidean global $AdS_{d+1}$ to
global $dS_{d+1}$ is
\be\label{EAdSglobaldSglobal}
ds^2 = L^2 (dr^2 + \sinh^2r\,d\Omega_d^2)
\quad\xrightarrow{\, L\ra -il,\ \ r\ra \tau-i\pi/2\, }\quad
ds^2 = l^2 (-d\tau^2 + \cosh^2\tau\,d\Omega_d^2)\,.
\ee
(see \eg\ \cite{Maldacena:2002vr}, \cite{Harlow:2011ke},
\cite{Maldacena:2019cbz}, for aspects of $AdS\leftrightarrow dS$
analytic continuations)\
A constant Euclidean time slice in Eucl $AdS$ is any $S^d$ equatorial
plane: this maps to a corresponding $S^d$ equatorial plane in $dS$.
The $AdS$ boundary at $r\ra\infty$ maps to the $dS$ future boundary
$I^+$ at $\tau\ra\infty$. Likewise, the analytic continuation from
Lorentzian global $AdS_{d+1}$ to $dS_{d+1}$ static coordinatization is
\be\label{dSstatAdS}
ds^2 = -\Big(1+{r^2\over L^2}\Big) dt^2 + {dr^2\over 1+{r^2\over L^2}}
+ r^2 d\Omega_{d-1}^2\ \ \xrightarrow{\, L\ra -il\,}\ \ 
ds^2 = -\Big(1-{r^2\over l^2}\Big) dt^2 + {dr^2\over 1-{r^2\over l^2}}
+ r^2 d\Omega_{d-1}^2\,.
\ee

Using this we can describe no-boundary de Sitter space (Lorentzian $dS$
in the top half joined smoothly at the midslice of time symmetry to a
Euclidean hemisphere in the bottom half) via analytic continuation from
$AdS$. In detail we have
\bea\label{nbdSstattoAdS}
&& ds^2_{(r>l)} = -{dr^2\over {r^2\over l^2}-1} + \Big({r^2\over l^2}-1\Big) dt^2
+ r^2 d\Omega_{d-1}^2 \quad \xrightarrow{\ l\ra iL\ } \nn\\ [-2mm]
&& \qquad\qquad\qquad\qquad\qquad\qquad\qquad\qquad
ds^2_{(r>L)} = -\Big(1+{r^2\over L^2}\Big) dt^2
+ {dr^2\over 1+{r^2\over L^2}} + r^2 d\Omega_{d-1}^2 \,,\nn\\ [1mm]
&& ds^2_{(r<l)} = \Big(1-{r^2\over l^2}\Big) dt_E^2 + {dr^2\over 1-{r^2\over l^2}}
+ r^2 d\Omega_{d-1}^2  \quad \xrightarrow{\ l\ra iL\ } \nn\\ [-2mm]
&& \qquad\qquad\qquad\qquad\qquad\qquad\qquad\qquad
ds^2_{(r<L)} = \Big(1+{r^2\over L^2}\Big) dt_E^2
+ {dr^2\over 1+{r^2\over L^2}} + r^2 d\Omega_{d-1}^2\,.\qquad
\eea
The above describes no-boundary $dS$ as an analytic continuation
from Lorentzian global $AdS$ ($r>L$) glued at $r=L$ with Euclidean
global $AdS$ ($r<L$).\
The $AdS$ boundary at $r\ra\infty$ maps to the $dS$ future boundary $I^+$ at
$r\ra\infty$, and the future universe $F$ parametrized by $r\in [\infty,l]$
maps to the $AdS$ region $r\in [\infty,L]$.
The $dS$ hemisphere is $\tau_E=-it=[0,{\pi\over 2}]$ where\
$-{dr^2\over {r^2\over l^2}-1}>0$\ is Euclidean (mapping from Euclidean
$AdS$).

\subsection{The IR extremal surface in\ $dS_{d+1}$}\label{sec:dSacIR}

Here we describe the IR extremal surfaces for maximal subregions in any
$dS_{d+1}$.\
The $AdS$ RT surfaces lie on some constant time slice: such a $t=const$
slice is 
\bea\label{nbdSstattoAdS2}
ds^2_{(r>l)} = -{dr^2\over {r^2\over l^2}-1} 
+ r^2 d\Omega_{d-1}^2 \ \ &\xrightarrow{\ l\ra iL\ }& \ \
ds^2_{(r>L)} = {dr^2\over 1+{r^2\over L^2}} + r^2 d\Omega_{d-1}^2 \,,\nn\\
ds^2_{(r<l)} =  {dr^2\over 1-{r^2\over l^2}}
+ r^2 d\Omega_{d-1}^2 \ \ &\xrightarrow{\ l\ra iL\ }& \ \
ds^2_{(r<L)} = {dr^2\over 1+{r^2\over L^2}} + r^2 d\Omega_{d-1}^2\,,
\eea
under the\ $dS\leftrightarrow AdS$\ analytic continuation.
Thus the $AdS$ side is entirely Euclidean on this slice so the
difference between the $r>L$ and $r<L$ parts in (\ref{nbdSstattoAdS})
disappears.
In the $dS$ static coordinatization, the future boundary $I^+$ is of
the form $R_t\times S^{d-1}_\Omega$ with $t, \Omega$ being spatial
coordinates. From the point of the Euclidean CFT$_d$ here, any of the
$S^{d-1}$ equatorial planes or the $t=const$ slice can be regarded as
boundary Euclidean time slices. So the analytic continuation from
$AdS$ leads to the $t=const$ surfaces as ``preferred'' boundary
Euclidean time slices, suggesting preferred observers at $I^+$.
These $t=const$ slices are the continuation to the future universe
of the constant time slices in the static patch (where $t$ is
Killing time).\
The global $dS_{d+1}$ metric (\ref{EAdSglobaldSglobal}), with $S^d$
cross-sections, restricted to any equatorial $S^d$ plane is identical
to the $t=const$ slice of the (Lorentzian) $dS$ static
coordinatization (\ref{dSstatAdS}), (\ref{nbdSstattoAdS2}), above:
\bea\label{dSglobaleqSd-dSstattconst}
&& ds^2_{global} 
\quad\xrightarrow{\,\theta_d=const\,}\quad
-d\tau^2+ l^2\cosh^2{\tau\over l}\, d\Omega_{d-1}^2\,, \nn\\
&& ds^2_{static} \quad\xrightarrow{\,t=const\,}\quad
-{dr^2\over {r^2\over l^2}-1} + r^2 d\Omega_{d-1}^2\,,
\qquad r = l\cosh{\tau\over l}\,.
\eea
Thus a generic equatorial plane in global $dS$ defines the same
boundary Euclidean time slice as the $t=const$ slice in the $dS$ static
coordinatization, and we will continue to use the parametrization
(\ref{nbdSstattoAdS2}).

The IR $AdS$ surface space spans the entire $AdS$ boundary
sphere. This continues to the IR $dS$ extremal surface (when the
subregion becomes the whole space at $I^+$), going from $r\ra\infty$
to $r=l$ as a timelike surface in the Lorentzian $dS$ region and then
in the hemisphere from $r=l$ to $r=r_*=0$ (where it turns around). The
turning point $r_*\in dS$ only exists in the Euclidean (hemisphere)
part of $dS$.

This IR surface is the boundary of the maximal subregion of the $S^{d-1}$
(\ie\ hemisphere) so it is anchored on the $S^{d-1}$ equator and wraps
the equatorial $S^{d-2}$\ (the red curve in Figure~\ref{fig1}). From
(\ref{nbdSstattoAdS}), (\ref{nbdSstattoAdS2}), it is clear that
the IR $dS$ extremal surface becomes a space-time rotation of that
in $AdS$.\ Its area continues as\ (with $R_c$ a cutoff at large $r$)
\bea\label{IRsurfAdSdS}
{V_{S^{d-2}}\over 4G_{d+1}} \int_0^{R_c} {r^{d-2} dr\over \sqrt{1+{r^2\over L^2}}}
\ & \xrightarrow{\, L\ra -il\,} &\
{V_{S^{d-2}}\over 4G_{d+1}} \int_0^l {r^{d-2} dr\over \sqrt{1-{r^2\over l^2}}}
+ {V_{S^{d-2}}\over 4G_{d+1}}
\int_l^{R_c} {r^{d-2} \sqrt{dr^2\over -({r^2\over l^2}-1)}} \ , \nn\\
&& =\ {1\over 2} {l^{d-1}V_{S^{d-1}}\over 4G_{d+1}} - i\# {l^{d-1}\over 4G_{d+1}}
{R_c^{d-2}\over l^{d-2}} +\, \ldots
\eea
where the $\ldots$ are subleading imaginary terms. The analytic
continuation helps map the area integrals but the $dS$ areas can
be straightforwardly evaluated the integrals in $dS$ form.
To see how the continuation works in detail, consider $AdS_4\ra dS_4$
and $AdS_3\ra dS_3$ above:
\be\label{IRsurfAdSdS4}
{V_{S^1}\over 4G_4}\int_0^{R_c}{rdr\over\sqrt{1+{r^2\over L^2}}} =
{\pi L^2\over 2G_4} \Big({R_c\over L} - 1\Big)\ \ \ra\ \
-i{\pi l^2\over 2G_4} {R_c\over l} + {\pi l^2\over 2G_4} = S^{IR}_{dS_4}\,,
\ee
and
\be\label{IRsurfAdSdS3}
{2L\over 4G_3}\log{R_c\over L}\ \ \ra\ \
-i{l\over 2G_3}\log{R_c\over l} + {\pi l\over 4G_3} = S^{IR}_{dS_3}\,.
\ee
For higher dimensions, there are further subleading imaginary terms in
the $dS$ area. These are of course the no-boundary extremal surface
areas (\ref{dS43rev}) found in \cite{Doi:2022iyj},
\cite{Narayan:2022afv}: we have simply tracked the analytic
continuation more closely here, while preserving the geometric picture
of space$\leftrightarrow$time rotation (in the spirit of
\cite{Narayan:2017xca,Narayan:2020nsc}), paving the way for further
analysis later.  This mapping between the $AdS$ and $dS$ extremal
surfaces can be done in detail for generic subregions as well but it
is most easily understood in the $dS_3$ case described later.

It is worth noting that the analytic continuations we have described
here are consistent with the geometric picture of $dS$ via space-time
rotation from $AdS$ (and so are slightly different from those in the
Poincare slicing in \cite{Narayan:2015vda,Narayan:2015oka,Sato:2015tta}
which involve imaginary time paths, although both are analytic
continuations of entanglement entropies in $AdS/CFT$).


\subsection{$dS_3 \leftrightarrow AdS_3$ extremal surfaces}\label{sec:AdS3dS3}

Here we describe the $dS_3$ case more elaborately here, for generic
non-maximal subregions which are straightforward to analyse explicitly
in $dS_3$. 
In this case the $AdS_3$ $t=const$ slice and the corresponding
$dS_3$ slice (\ref{nbdSstattoAdS2}) are
\be\label{AdS3tslicedS3}
ds^2 =  {dr^2\over 1+{r^2\over L^2}} + r^2 d\theta^2\ \
\xrightarrow{\, L\ra -il\,}\ \ 
ds^2 = - {dr^2\over {r^2\over l^2}-1} + r^2 d\theta^2\,.
\ee
The area functionals become
\be\label{AdSdSareaFnal}
{2\over 4G_3} \int_{R_c}^{r_*}
dr\,\sqrt{{1\over 1+{r^2\over L^2}} + r^2 (\theta')^2}
\ \ \xrightarrow{\, L\ra -il\,}\ \
{2\over 4G_3} \int_{R_c}^{r_*} dr\,\sqrt{ -{1\over  {r^2\over l^2}-1}
+ r^2 (\theta')^2}\ .
\ee
We have written the $dS$ integrand to illustrate the timelike nature
for $r>l$ in the Lorentzian part while for $r<l$ the expression under
the radical is Euclidean and positive.\
The extremization leads to a conserved quantity
${r^2\theta'\over\sqrt{...}}=A$\,: this leads to
\bea
(\theta')^2 = {1/r^2\over (1+{r^2\over L^2})({r^2\over A^2}-1)}\,,\ &&\
S = {2\over 4G_3} \int_{R_c}^{r_*} dr\,\sqrt{{1\over 1+{r^2\over L^2}}
  \Big({1\over 1-{A^2\over r^2}}\Big)}\  \qquad [AdS]\,; \nn\\
(\theta')^2 = {1/r^2\over (1-{r^2\over l^2})({r^2\over A^2}-1)}\,,\ &&\
S = {2\over 4G_3} \int_{r_*}^{l} dr\,\sqrt{{1\over 1-{r^2\over l^2}}
  \Big({1\over 1-{A^2\over r^2}}\Big)}\ . \qquad [dS:\ r<l]\,,\nn\\
(\theta')^2 = {1/r^2\over ({r^2\over l^2}-1)(1+{r^2\over A^2})}\,,\ &&\
S = {-2i\over 4G_3} \int_{R_c}^{r_*} dr\,\sqrt{{1\over {r^2\over l^2}-1}
  \Big({1\over 1+{A^2\over r^2}}\Big)}\ . \qquad [dS:\ r>l]\,.\ \ \
\eea
These are related by the analytic continuation\ $L\ra -il,\ A^2\ra A^2$\
for $AdS$ to $dS,\ r<l$ (hemisphere) and\ $L\ra -il,\ A^2\ra -A^2$\
for $AdS$ to $dS,\ r>l$ (Lorentzian). In the Lorentzian $dS$ region,
the second mapping (on $A^2$) is necessary since there is extra data
pertaining to the extremal surface (in particular the turning point)
which also needs to be continued appropriately: as $r\ra\infty$ we 
have $(\theta')^2\ra {A^2\over l^2r^6}$ so $(\theta')^2>0$ is
well-defined near $I^+$. Instead had we retained $A^2$ as it is from
$AdS$, we would have obtained\
$(\theta')^2 = {1/r^2\over ({r^2\over l^2}-1)(1-{r^2\over A^2})}$ so
$(\theta')^2<0$ when $r>A$ (even for infinitesimal $A^2=\varepsilon^2>0$).
In the top Lorentzian part we have $r>l$ so $(\theta')^2<0$ is
ill-defined as a curve in the Lorentzian $dS$ (recalling
\cite{Narayan:2015vda}): continuing $A^2\ra -A^2$
allows interpretation as a timelike surface in the Lorentzian part
of $dS$, which then glues onto the spatial surface in the hemisphere.

\begin{figure}[h] 
\hspace{-0.5pc}
\includegraphics[width=18pc]{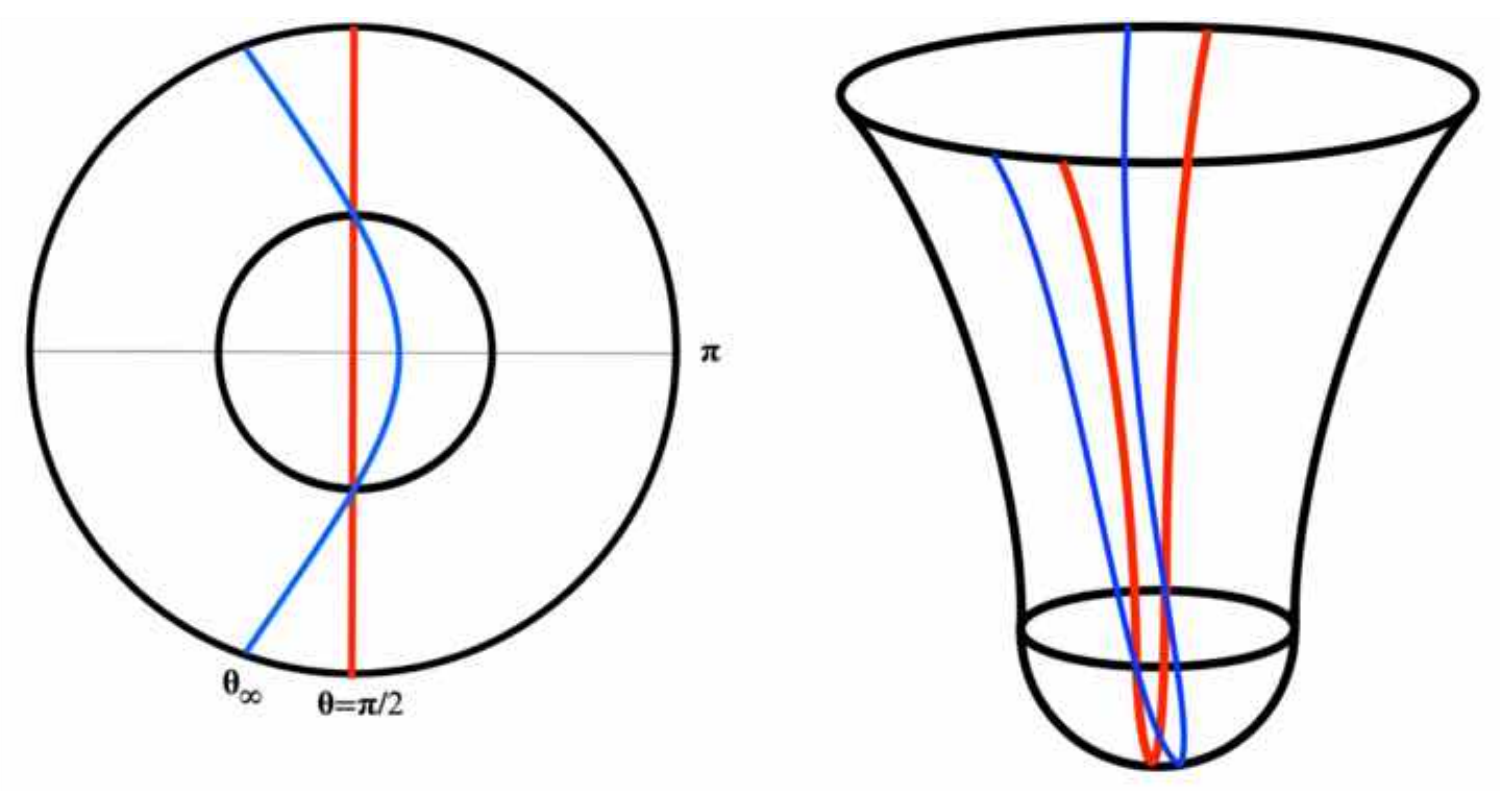}
\hspace{0.5pc}
\begin{minipage}[b]{19pc}
\caption{{ \label{fig1}
    \footnotesize{
        $dS$ no-boundary extremal surfaces in the $t=const$ slice in
        the static coordinates (right side picture;\ left side is the
        ``top view'' from $I^+$). The red curve is the IR extremal
        surface for maximal boundary subregion
        $[-{\pi\over 2}, {\pi\over 2}]$\ \ (hemisphere in $S^{d-1}$).
        The blue curve is the extremal surface when the subregion
        $[-\theta_\infty, \theta_\infty]$ at $I^+$ is not maximal.
        }}}
\end{minipage}
\end{figure}

These can now be solved for $\theta(r)$ explicitly giving
\bea
(\theta')^2 = {1/r^2\over (1+{r^2\over L^2})({r^2\over A^2}-1)}\,,
&& \  \tan(\theta) =  
           {\sqrt{1+{r^2\over L^2}}\over \sqrt{{r^2\over A^2}-1}}\,,
\qquad\qquad\quad [AdS] \nn\\
(\theta')^2 = {1/r^2\over (1-{r^2\over l^2})({r^2\over A^2}-1)}\,,
&& \ \tan\big(\theta-{\pi\over 2}\big)
= {\sqrt{1-{r^2\over l^2}}\over\sqrt{{r^2\over A^2}-1}}\,,
\qquad\quad  [dS: r<l] \nn\\  
(\theta')^2 = {1/r^2\over ({r^2\over l^2}-1)(1+{r^2\over A^2})}\,,
&& \ \tan\big(\theta-{\pi\over 2}\big)
= -{\sqrt{{r^2\over l^2}-1}\over \sqrt{{r^2\over A^2}+1}}\,,
\qquad\ [dS: r>l].\quad    
\eea
As is clear, these are again related via the analytic continuations
we have stated above, \ie\ $L\ra -il,\ A^2\ra A^2$\ for $AdS$ to
$dS,\ r<l$ (hemisphere) and\ $L\ra -il,\ A^2\ra -A^2$\ for $AdS$ to
$dS,\ r>l$ (Lorentzian).
The asymptotics in the $dS$ case are
\be\label{dS3asympAtheta_infty}
\theta\xrightarrow{r\ra\infty} {\pi\over 2} - \tan^{-1}{A\over l}
\equiv \theta_\infty\,;\qquad \theta\xrightarrow{r\ra l} {\pi\over 2}\,;
\qquad \theta\xrightarrow{r\ra A} \pi\,.
\ee
Thus this describes a surface anchored at $\theta_\infty\in I^+$ going
into the time direction as a timelike surface, hitting the $r=l$ slice
at $\theta={\pi\over 2}$ and then going around the hemisphere till
$\theta=\pi$ at the turning point $r=A$. This then joins with a similar
half-surface on the other side (of $\theta=0$, so $\theta<0$ now) going
from $\theta=\pi$ at $r=A$ to $\theta={\pi\over 2}$ at $r=l$ and thence
to $-\theta_\infty\in I^+$. The boundary subregion at $I^+$ is
$[-\theta_\infty,0]\cup[0,\theta_\infty]$, with $\theta_\infty$ defined
by the parameter $A^2>0$ as above.
The parametrization above has a symmetry about $\theta=0$ and picks
out $\theta=\pm {\pi\over 2}$ as the point where the surface hits the
$r=l$ slice: translating in $\theta$ modifies the parametrization
above to describe surfaces with different anchoring points and $r=l$
points. The overall picture is shown in Figure~\ref{fig1}.

For $A=0$ which is the IR surface the above give\ $(\theta')^2=0$
and the surface lies on the $\theta={\pi\over 2}$ subplane, in the
above parametrization (or equivalently any $\theta=\theta_\infty$).
The subregion at $I^+$ is now maximal, being half the circle
(parametrized as $[-{\pi\over 2},0]\cup[0,{\pi\over 2}]$).
The area of this IR surface can be evaluated setting $d=2$ in
(\ref{IRsurfAdSdS}) and gives (\ref{IRsurfAdSdS3}).

For $A$ small, we expect the $dS$ extremal surface to be a small
perturbation around the IR surface: indeed this can be seen to be
true, with the surface described as
\be\label{dS3smallA}
\tan\big(\theta-{\pi\over 2}\big)
= {\sqrt{1-{r^2\over l^2}}\over\sqrt{{r^2\over A^2}-1}}\,,
\quad  [r<l];\qquad
\ \tan\big(\theta-{\pi\over 2}\big)
= -{A\over l} \sqrt{1-{l^2\over r^2}}\,, \quad [r>l].
\ee  

The turning point is in the hemisphere region $r<l$ in $dS$. To see
this, note that perturbing away from the IR surface $A=0$ by increasing
$A$ moves the turning point to $r=A$ where $(\theta')^2\ra\infty$.
However $(\theta')^2>0$ gives $r>A$. This can also be seen by
parametrizing $r=l\sin\psi$ so $r=l$ is $\psi={\pi\over 2}$: then
$({d\theta\over d\psi})^2={A^2/\sin^2\psi\over l^2\sin^2\psi-A^2}$
so $A^2<l^2$. In the surface equation, the limit $A\ra l$ gives
$\tan\theta\ra\sqrt{{r^2/l^2-1\over 1-r^2/l^2}}=\pm i$, so $\theta$
stops being a well-defined real angle variable. Thus we have
$0\leq A^2<l^2$, and then (\ref{dS3asympAtheta_infty}) implies that
$|\theta_\infty|\geq {\pi\over 4}$.

The area of this extremal surface for general subregion (general $A$)
has contributions from the top timelike part and the spatial hemisphere
part. In the Euclidean hemisphere part of $dS$, we obtain
\be
S_t^{r<l} = \int_{l}^{A} {2dr\over 4G_3} \sqrt{{1\over 1-{r^2\over l^2}}
  \Big({1\over 1-{A^2\over r^2}}\Big)}
= {l\over 2G_3}\sin^{-1}\sqrt{{r^2-A^2\over l^2-A^2}}\,\Big|_A^l
= {\pi l\over 4G_3}\,,
\ee
and the Lorentzian part of $dS$ gives 
\bea
S_t^{r>l} &=& -i\int_{R_c}^{l} {2dr\over 4G_3} \sqrt{{1\over {r^2\over l^2}-1}
  \Big({1\over 1+{A^2\over r^2}}\Big)}
= i{l\over 2G_3}\log(\sqrt{r^2+A^2}-\sqrt{r^2-l^2})|_l^{R_c} \nn\\
&=& -i{l\over 2G_3}\log{R_c\over l} -i{l\over 4G_3}\log(\sin^2\theta_\infty)\,,
\eea
where we have expanded to get the leading divergence and the finite
part, using (\ref{dS3asympAtheta_infty}). So
\be\label{SdS3-gentheta}
S_t=S_t^{r>l}+S_t^{r<l} = -i{l\over 2G_3}\log{R_c\over l}
- i{l\over 4G_3}\log(\sin^2\theta_\infty) + {\pi l\over 4G_3}\,,
\ee
in agreement with \cite{Hikida:2022ltr}.
The IR limit $A=0$ gives $\theta_\infty={\pi\over 2}$ and
the area (\ref{SdS3-gentheta}) becomes (\ref{IRsurfAdSdS3}).

\subsection{Higher dimensions, $dS_{d+1}$}\label{sec:dSd+1}

It is convenient to use\
$ds^2 = {l^2\over\tau^2} \big(-{d\tau^2\over 1-\tau^2} + (1-\tau^2) dw^2
+ d\Omega_{d-1}^2 \big)$, where $\tau={l\over r} ,\ \ w={t\over l}$,\
to describe the $dS_{d+1}$ static coordinatization.
In this case, the area functional on the $w=const$ or $t=const$ slice
in the top Lorentzian part (analogous to (\ref{AdSdSareaFnal}) for
$dS_3$) is \cite{Narayan:2017xca,Narayan:2020nsc}
\be
S = -i\, {l^{d-1} V_{S^{d-2}}\over 4G_{d+1}}
\int {d\tau\over \tau^{d-1}} (\sin\theta)^{d-2}
\sqrt{{1\over 1-\tau^2} - (\theta')^2}\ .
\ee
This also applies to any $S^d$ equatorial plane in global $dS$ (from
(\ref{dSglobaleqSd-dSstattconst})).
The equation of motion\ ${d\over d\tau} ({\del L\over\del \theta'}) =
{\del L\over\del\theta}$\ becomes
\be\label{tconstEOM}
   {d\over d\tau} \left({-\theta'\over\sqrt{{1\over 1-\tau^2}-(\theta')^2}}
   {(\sin\theta)^{d-2}\over \tau^{d-1}}\right) =
   (d-2) {(\sin\theta)^{d-3}\over \tau^{d-1}}
   \cos\theta \sqrt{{1\over 1-\tau^2}-(\theta')^2}\ ,
\ee
which in general is difficult to solve exactly. However it is
straightforward to realize the IR extremal surface at
$\theta={\pi\over 2}$ where the right side vanishes (and $\theta'=0$).

A small perturbation about this IR surface is parametrized as\
$\theta(\tau)={\pi\over 2}-\delta\theta(\tau)$. Then
$\cos\theta\sim \delta\theta(\tau)$ to $O(\delta\theta)$ giving
the linearized equation 
\be
   {d\over d\tau} \left({\sqrt{1-\tau^2}\over \tau^{d-1}}\
   {d\delta\theta(\tau)\over d\tau}\right) =
   {d-2\over \tau^{d-1}} {\delta\theta(\tau)\over\sqrt{1-\tau^2}}
   \quad\ra\quad \delta\theta(\tau)={A\over l}\sqrt{1-\tau^2}\ .
\ee
This solution can be seen to satisfy the linearized equation for
$dS_{d+1}$ in any dimension. Of the two solutions, we have picked the
one that has regularity properties and satisfies the boundary
condition $\theta={\pi\over 2}-{A\over l}$ at $I^+$\ ($r\ra\infty$).
This finally gives the near-IR surface
\be
\theta(r)={\pi\over 2}-{A\over l}\sqrt{1-{l^2\over r^2}}\,;\qquad
\theta\xrightarrow{r\ra\infty} {\pi\over 2}-{A\over l}\equiv
\theta_\infty\,,\quad \theta\xrightarrow{r\ra l} {\pi\over 2}\,.
\ee
Thus for small $A$, we finally have the same form of $\theta(r)$ as
in (\ref{dS3smallA}) for $dS_3$ in the top Lorentzian part.

For the hemisphere part, the extremal surface area functional is
\be
S = {l^{d-1} V_{S^{d-2}}\over 4G_{d+1}}
\int dr\, r^{d-2} (\sin\theta)^{d-2}
\sqrt{{1\over 1-{r^2\over l^2}} + r^2(\theta')^2}\,,
\ee
and the equation of motion can be seen to show the IR surface
solution at $\theta={\pi\over 2}$ and $\theta'=0$.
Near the $r=l$ slice, the near-IR surface can be parametrized as\
$\theta(\tau)={\pi\over 2}-\delta\theta(\tau)$\ giving
\be
   {d\over dr} \left(r^d\,\sqrt{1-{r^2\over l^2}}\
     {d\delta\theta(r)\over dr} \right)
= -{(d-2)\, r^{d-2}\, \delta\theta(r)\over\sqrt{1-{r^2\over l^2}}}
\ee
which shows\ $\delta\theta(r)\sim -{A\sqrt{1-r^2/l^2}\over r}$ which
is similar in form to the $dS_3$ case for small $A$, and near $r=l$,
where it glues onto the top timelike surface.
As the surface dips in further, it is not a small deviation from
$\theta={\pi\over 2}$ and eventually hits $\theta=\pi$. Perturbing
near $\theta=\pi$ suggests a solution smooth in the vicinity and
turning around. This confirms a consistent picture for the
bottom hemisphere great-circle-like extremal surface stretching
along the $\theta$-direction between $\theta={\pi\over 2}$ and $\pi$
(and wrapping the other angular directions) similar to the $dS_3$
case resulting in the picture in Figure~\ref{fig1}., with area half
de Sitter entropy.

\subsection{Subregion duality, geometrically}\label{sec:subregDuality}

There is a simple way to visualize the geometric picture of these
$dS$ extremal surfaces, which are geodesics in $dS_3$: we know that
all geodesics in the hemisphere are parts of great circles. Any great
circle hits the equator at antipodal points, say $\theta=\pm{\pi\over 2}$.
The great circle corresponding to the IR surface $A=0$ is ``vertical'':
it joins to timelike curves that go ``vertically up'' so
$\theta=\pm{\pi\over 2}$ for the two half-surfaces on either side
(of $\theta=0$). As we now increase $A$, the great circle in the
hemisphere ``tilts'', still anchored at $\theta=\pm{\pi\over 2}$:
the timelike curves in the Lorentzian part now correspondingly tilt
to hit $I^+$ at some $|\theta_\infty|<{\pi\over 2}$. Roughly the IR
surface has been perturbed to now acquire an overall tilt, with
$\theta$ going from $\theta_\infty$ to ${\pi\over 2}$ to $\pi$ at $r=A$.
This is depicted in Figure~\ref{fig1}, the right side figure being
no-boundary $dS$ with the left side figure being the ``top view''
from $I^+$. The outer circle is at $I^+$, while the inner black circle
is the $r=l$ plane where the hemisphere glues on. The red curve is the
``vertical'' IR extremal surface anchored at the boundary of the
maximal subregion (semicircle arc) while the blue one is anchored at
the boundary of the smaller boundary subregion arc, and gives the
tilted curve in the right side figure.

\begin{figure}[h] 
\hspace{2pc}
\includegraphics[width=33pc]{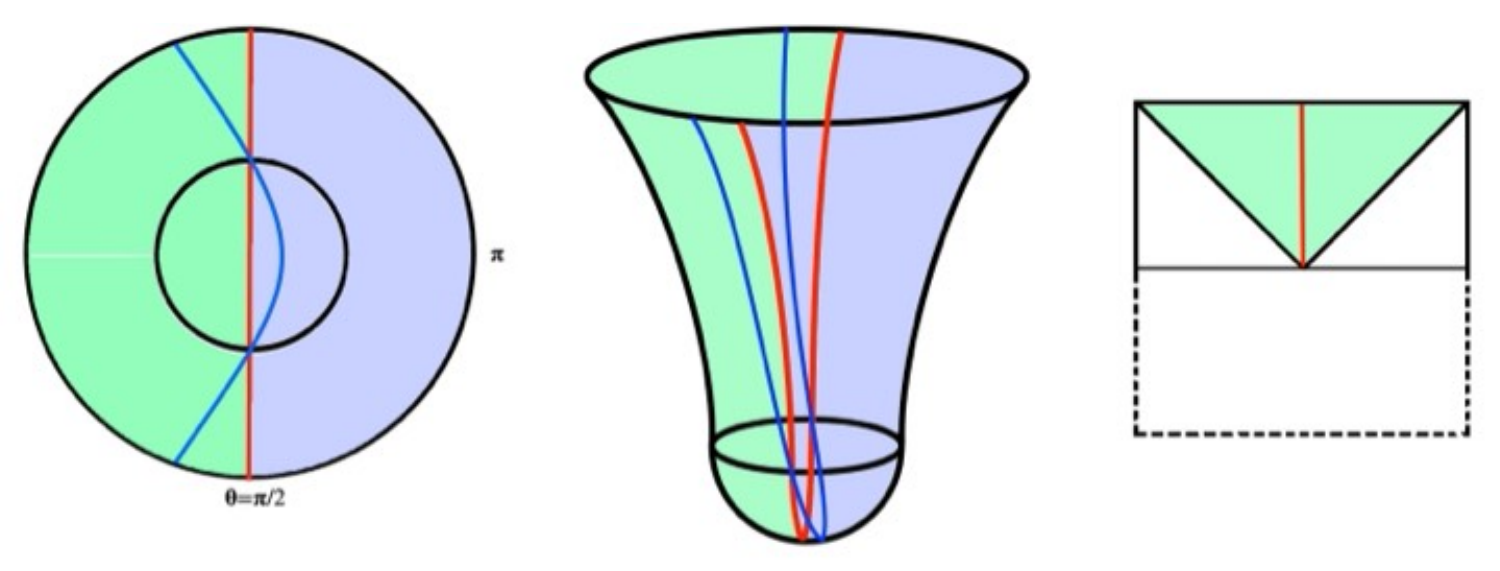}
\caption{{ \label{fig2}
    \footnotesize{
      $dS$ no-boundary extremal surfaces in the $t=const$ slice
      (middle picture;\ left side is the ``top view'' from $I^+$). The
      red curve is the extremal surface for maximal boundary
      subregion, and the blue curve when the subregion is smaller.
      The green and violet shaded regions are the bulk subregions
      defined by the time-entanglement or pseudo-entanglement wedges
      restricted to this $t=const$ slice.  The right side figure is
      the bulk region in the top Lorentzian half in the full $dS$
      Penrose diagram, including the $t$ direction: the IR surface
      lies on the $t=const$ slice which is depicted as the red
      vertical line.        }}}
\end{figure}
This leads to a heuristic version of subregion-subregion duality in
$dS$, but with some new features relative to the $AdS$ case
\cite{Czech:2012bh,Wall:2012uf,Headrick:2014cta} (and the reviews
\cite{Rangamani:2016dms}, \cite{Harlow:2018fse}, \cite{Headrick:2019eth}).
For instance in Figure~\ref{fig2}, the green shaded region is the
bulk subregion or
the ``time-entanglement wedge'' or the ``pseudo-entanglement wedge''
(restricted to this $t=const$ slice) corresponding to the maximal
boundary subregion: we are defining this, just geometrically, as the
bulk region bounded by the IR extremal surface and the boundary
subregion. Likewise the other maximal (semicircular) subregion gives
the complementary violet shaded bulk subregion. In Figure~\ref{fig2},
the left and middle pictures pertain to the $t=const$ boundary
Euclidean time slice: to obtain the full bulk time-entanglement or
pseudo-entanglement wedge, we need to consider the full de Sitter
space, including the $t$ direction.  This is shown in the right
picture: $t=const$ slices are straight lines from $I^+$ to the
bifurcation point in the top Lorentzian half in the Penrose
diagram (the bottom half, with dashed lines, is replaced by the
Euclidean hemisphere). The IR surface on the $t=const$ slice is
shown as the vertical red line. To define the time-entanglement or
pseudo-entanglement wedge, first note that from the bottommost point
of the extremal surface (red curve) we can draw out four spacetime
regions: the left and right wedges, and the top and bottom wedges. It
appears consistent to define the pseudo-entanglement wedge as the top
wedge, bounded by the boundary subregion at $I^+$. In some sense,
this is the domain of dependence of the analog of the $AdS$ homology
surface here. This can also be seen to arise from the analytic
continuation of the familiar entanglement wedge in $AdS$: given that
the analytic continuation (\ref{nbdSstattoAdS}),
(\ref{nbdSstattoAdS2}), (\ref{IRsurfAdSdS}), of the slice and the IR
surface amount to a space-time rotation from $AdS$ to $dS$, it is not
surprising that the pseudo-entanglement wedge resembles a space-time
rotation from the entanglement wedge in $AdS$. Finally, we see that
this top (green) wedge is the future universe in the $dS$ static
coordinatization.

It is important to note that our discussions here are simply
geometric, in analogy with the $AdS$ case, and fuelled by the analytic
continuation. Along these lines and using the causality arguments in
\cite{Headrick:2014cta}, the bulk pseudo-entanglement wedge for a
generic subregion at the future boundary $I^+$ would be expected to
not lie within its causal wedge which is the past lightcone wedge (the
past domain of dependence). The extremal surfaces extend in timelike
directions away from the subregion, suggesting causal consistency (see
also \cite{Narayan:2020nsc}).  One might expect a bulk reconstruction
relation of the form\ $\phi(X)=\int_{I^+} d^d\sigma G(X,\sigma)
O(\sigma)$ with $O(\sigma)$ defined on $I^+$ while $\phi(X)$ contains
bulk time as well. This is analogous to bulk field expansions
$\phi(\sigma,\tau)\sim \int_{{\vec k}} G({\vec k},\tau,\sigma)
\phi_0({\vec k})$ employed in the late-time Wavefunction in $dS/CFT$
\cite{Maldacena:2002vr}, with $\phi_0({\vec k})$ the future boundary
conditions at $I^+$.  It would be interesting to understand more
directly if these heuristic geometric observations and the associated
(pseudo-)entanglement wedge reconstruction can be ``derived'' from
analogs in $dS/CFT$ of modular flow, relative entropy, error
correction codes and so on
\cite{Almheiri:2014lwa,Jafferis:2015del,Dong:2016eik} (see also the
reviews \cite{Rangamani:2016dms}, \cite{Harlow:2018fse},
\cite{Headrick:2019eth}).


\begin{figure}[h] 
\hspace{0.25pc}
\includegraphics[width=11pc]{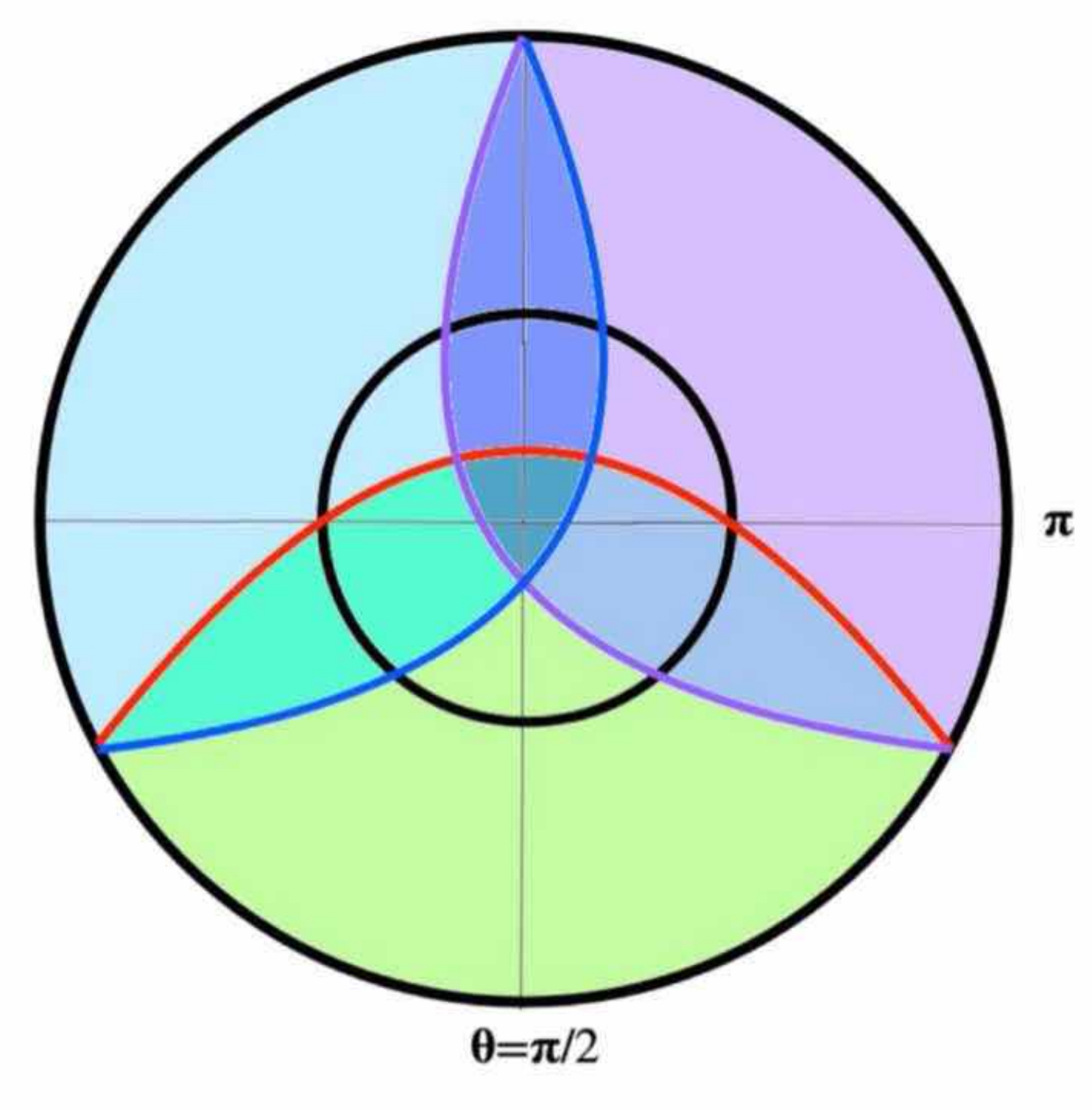}
\hspace{0.5pc}
\begin{minipage}[b]{26pc}
\caption{{ \label{fig3}
    \footnotesize{
      $dS$ no-boundary extremal surfaces for disjoint subregions
      at $I^+$ in the $t=const$ slice (static coordinatization): this 
      is the ``top view'' from $I^+$ . These are the red, violet,
      and blue curves for the three boundary subregions at $I^+$. The
      green, violet and blue shaded regions are the corresponding bulk
      subregions. The bulk subregions overlap (represented by the
      different color shadings in the various overlaps; \eg\ light
      blue overlapping with violet leads to darker blue-violet etc).
      \newline      }}}
\end{minipage}
\end{figure}


For these maximal boundary subregions which are disjoint at
$I^+$\ (the two half-circles), the bulk subregions are also disjoint
with no overlap: Figure~\ref{fig2} shows that the green and violet
bulk subregions are disjoint (separated by the IR extremal surface
(red)). Thus these maximal subregions satisfy subregion duality
geometrically, \ie\ disjoint boundary subregions are dual to disjoint
bulk subregions. However generic non-maximal boundary subregions
do not appear to satisfy this (naive) subregion duality, suggesting
the only well-defined subregions are maximal ones. (Similar issues
may arise for timelike intervals in $AdS/CFT$.)

To elaborate,
consider Figure~\ref{fig3}, depicting the ``top view'' of three
disjoint generic boundary subregions at $I^+$ defined by the arcs bounded by
the red, violet and blue curves.  The red curve is manifestly seen to
be symmetric about the $\theta={\pi\over 2}$ equatorial line: it
intersects the $r=l$ inner circle at the $\theta=\pi$ equatorial line
(normal to the $\theta={\pi\over 2}$ line). Generic extremal surface
curves for other subregions can be drawn likewise by rotating the red
curve about the origin of the circles (so they intersect the $r=l$
inner circle at their corresponding ${\tilde\theta}={\pi\over 2}$
equatorial line): thus we obtain the violet and blue curves. Now it is
clear that the bulk subregions overlap and are not disjoint, even
though the corresponding boundary subregions are disjoint. The
detailed overlaps can be seen by labelling the green, violet, blue
regions as $g, v, b$: then $b+v$ is the blue-violetish overlapping
region (colored darker blue-violet) etc, while the central triangular
(dark turquoise) region is $g+b+v$. This is the ``top view'' here
(suppressing the analog of the middle image in Figure~\ref{fig2}), so
it is somewhat distinct from the corresponding picture in $AdS$ of
disjoint subregions on a constant time slice.

\begin{figure}[h] 
\hspace{0.5pc}
\includegraphics[width=9pc]{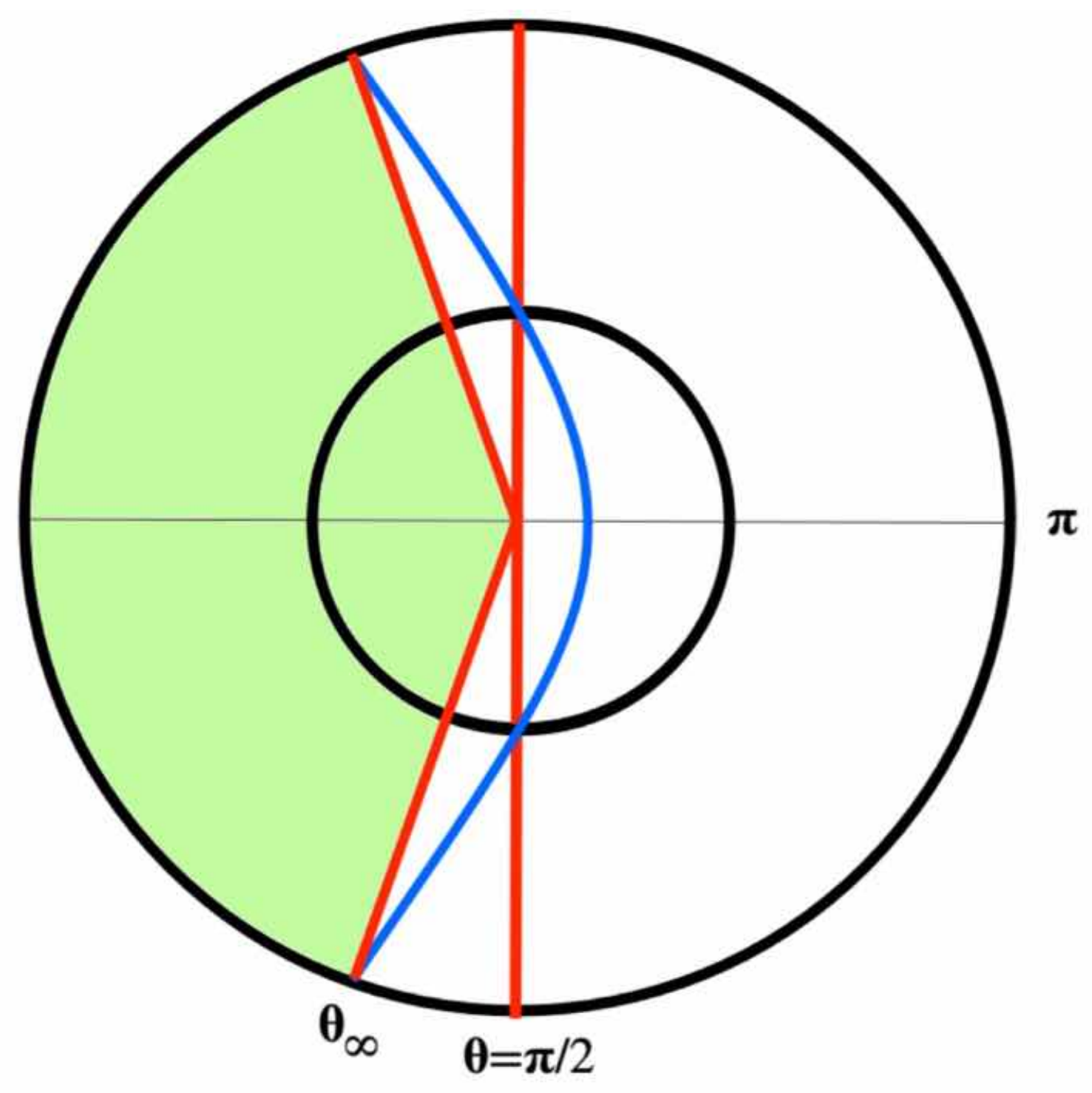}
\hspace{0.5pc}
\begin{minipage}[b]{25pc}
\caption{{ \label{fig4}
    \footnotesize{
      $dS$ no-boundary extremal surfaces and alternative ways to
      define bulk subregions (``top view'' from $I^+$). The bulk
      subregions for disjoint boundary subregions at $I^+$ do not
      overlap now: however the extremal surface now has a cusp at
      the no-boundary point (the earlier smooth extremal surface is
      the blue curve).
      \newline      }}}
\end{minipage}
\end{figure}
There is an alternative way to define bulk subregions
geometrically, which leads to the picture in Figure~\ref{fig4}.
Consider the boundary subregion $[-\theta_\infty,\theta_\infty]$.
We construct the extremal surface through the intersections of
the maximal (red) extremal surfaces for
$[-\theta_\infty,-\theta_\infty+\pi]$ and for 
$[\theta_\infty-\pi,\theta_\infty]$: this leads to the two
red half surfaces which meet at the origin, \ie\ the no-boundary
point in the center, but with a cusp since this effectively
involves the intersection of two great circles\ (by comparison
the earlier extremal surface is smooth at the no-boundary point).
The green bulk subregion is obtained as the region bounded
by this cuspy extremal surface and the boundary subregion.
Perhaps such cuspy extremal surfaces should be discarded,
favouring the earlier smooth ones.


\bigskip

\noindent {\bf \emph{Entirely Lorentzian $dS$, future-past surfaces}}
\vspace{2mm}

\begin{figure}[h] 
\hspace{1pc}
\includegraphics[width=13pc]{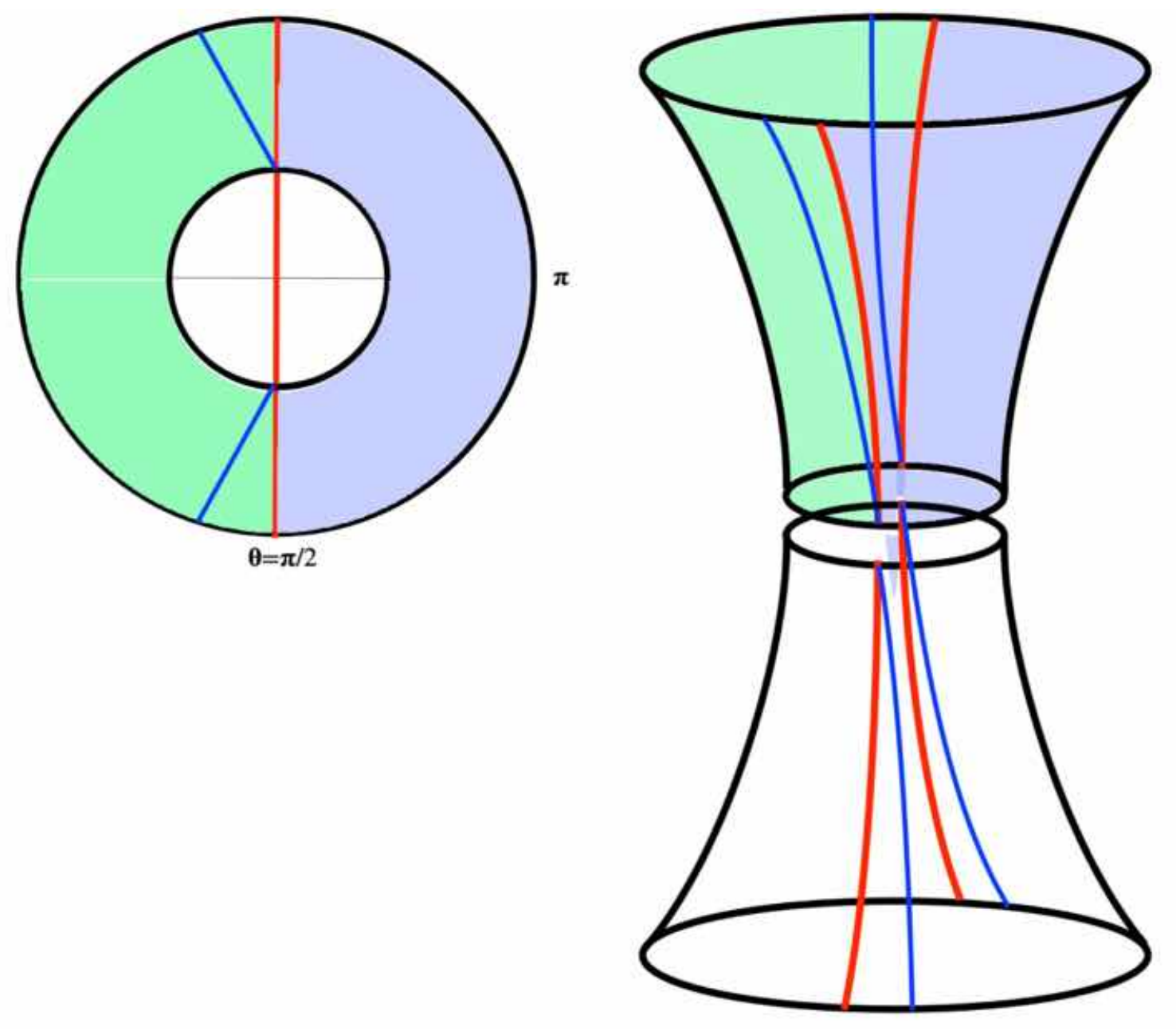}
\hspace{2pc}
\begin{minipage}[b]{20pc}
\caption{{ \label{fig5}
    \footnotesize{
      $dS$ future-past extremal surfaces from two copies of no-boundary
      surfaces (Figure~\ref{fig2}). The two copies of no-boundary $dS$
      are glued at the $r=l$ plane (removing the bottom hemisphere).
      \newline\newline     }}}
\end{minipage}
\end{figure}
So far we have been describing no-boundary de Sitter extremal
surfaces.  To discuss future-past surfaces, we first construct
entirely Lorentzian de Sitter space by gluing two copies of
no-boundary de Sitter space as in Figure~\ref{fig5}, with a top
Lorentzian half glued onto a bottom Lorentzian half smoothly at the
$r=l$ midslice, removing the Euclidean hemispheres in both.  We now
construct future-past surfaces in entirely Lorentzian de Sitter by
gluing two copies of the no-boundary surfaces. For instance, the top
timelike part of the (red) IR surface for maximal $I^+$ subregion
glues smoothly at $r=l$ onto the bottom timelike IR surface anchored
at $I^-$, to give the entirely timelike IR future-past surface
stretching between $I^\pm$. The corresponding time-entanglement or
pseudo-entanglement wedge comprises the future and past universe,
\ie\ the rightmost image in Figure~\ref{fig2} (the future universe
in the top Lorentzian $dS$) and its past copy. The blue curve shows
the future-past extremal surface for non-maximal $I^\pm$ subregions.
It is worth noting though that reconstructing the entirely Lorentzian
bulk appears to involve the two copies in a nontrivial manner (encoding
$|\Psi_{dS}|^2$ alongwith the area relations (\ref{dS43rev2})).

Finally, it is worth noting that our geometric discussions of the
time-entanglement wedge, or pseudo-entanglement wedge, are based on
the extremal surfaces on the $t=const$ slice we have been studying,
which can be obtained from the analytic continuations of constant time
slice RT surfaces in $AdS$. It can be seen that the resulting pictures
are somewhat different from those in \cite{Narayan:2020nsc} which
discussed analogs of the entanglement wedge based on future-past
extremal surfaces on $S^d$ equatorial planes in the $dS$ static
coordinatization (which would map to equatorial plane slices in
$AdS$). These differences are not surprising since the
corresponding boundary Euclidean time slices are quite different.
The analytic continuation from $AdS$ suggests the slices here as
``preferred'' as we have stated previously: it would be interesting
to understand this better.

\subsection{Entropy relations and inequalities}\label{sec:MITISSB}

The $AdS$ RT/HRT surface areas are known to strikingly satisfy various
entanglement entropy inequalities: we now study $dS$ analogs of some
of these using illustrative examples.  First, for the IR
subregion with $\theta_\infty={\pi\over 2}$\,, we
have\ $S[A_{IR}]=S[A_{IR}^C]$ from (\ref{SdS3-gentheta}), \ie\ the
time-entanglement for the IR subregion and its complement (also an IR
subregion) are equal: on geometric grounds it is consistent to take
$S[A_{IR}^C]=S[A_{IR}]$ for generic $\theta_\infty$ subregions and we
will use this in what follows.  Since the complex-valued areas are
somewhat different from the familiar ones in $AdS$, it is unclear if
$S[A_{IR}^C]=S[A_{IR}]$ implies purity rigorously.

Using the area expressions in $dS_3$ (\ref{SdS3-gentheta}), we can
evaluate formal analogs of mutual information for disjoint boundary
subregions. Consider two disjoint adjacent subregions $A, B$, with
$|\theta_\infty|={\pi\over 4}$ each (so each is a quadrant with spread
$2\theta_\infty={\pi\over 2}$;\ see Figure~\ref{fig3}): the union
subregion $A\cup B$ is
the single contiguous $IR$ subregion with $\theta_\infty={\pi\over 2}$
(since $A, B$ are adjacent): then, using (\ref{SdS3-gentheta}),
we have the ``mutual time-information'' or ``mutual pseudo-information''
\bea\label{MI-t}
I_t[A,B] &=& S[A]+S[B] - S[A\cup B] \nn\\
&=& 2\Big(-i{l\over 2G_3}\log{R_c\over l}
- i{l\over 4G_3}\log(\sin^2{\pi\over 4}) + {\pi l\over 4G_3}\Big)
+ i{l\over 2G_3}\log{R_c\over l} - {\pi l\over 4G_3} \nn\\
&=& - i{l\over 2G_3}\log{R_c\over l} + i{l\over 2G_3}\log 2
+ {\pi l\over 4G_3}\,.\nn\\
\Rightarrow && {\rm Re}\, I_t[A,B] > 0\,,\qquad {\rm Im}\, I_t[A,B]<0\,.
\eea
This is also the pattern for adjacent disjoint subregions with
other $\theta_\infty$ values, as can be seen using the area expression
(\ref{SdS3-gentheta}), which gives\
${\rm Re} I_t[A,B] > 0, \ {\rm Im} I_t[A,B]<0$.

Now let us consider three disjoint adjacent subregions $A, B, C$, each
a quadrant with $|\theta_\infty|={\pi\over 4}$ each. Together they cover
three quadrants of the full circle at $I^+$\ (schematically picking three
of the four quadrants in Figure~\ref{fig3}). Recalling \cite{Hayden:2011ag}
in the $AdS$ case, we want to consider the time-entanglement analogs
of the inequalities associated with strong subaddivity and tripartite
information, using the $dS_3$ areas (\ref{SdS3-gentheta}). Then
$AB\equiv A\cup B$ and $BC\equiv B\cup C$ are maximal subregions with
spread $2\theta_\infty=\pi$ each (with area given by the IR surface area
(\ref{IRsurfAdSdS3})), while $AC\equiv A\cup C$ comprises antipodal
quadrants with the extremal surface comprising an ``inner'' component
with area $S_{A_+,C_-}$ and an ``outer'' one with area $S_{A_-,C_+}$\
(where $A$ and $C$ have endpoints $A_-,A_+$ and $C_-,C_+$). These
are equivalent respectively to extremal surfaces for $B$ and its
antipodal quadrant. Finally $ABC\equiv A\cup B\cup C$ has area given
by its complementary quadrant. Putting these together, we have
\bea\label{SABC-pi/4}
S_A = S_B = S_C = -i{l\over 2G_3}\log{R_c\over l}
- i{l\over 4G_3}\log(\sin^2{\pi\over 4}) + {\pi l\over 4G_3}
\equiv S_{\pi/4}\,,\qquad\quad && \nn\\
S_{AB} = S_{BC} = -i{l\over 2G_3}\log{R_c\over l} + {\pi l\over 4G_3}
\equiv S_{\pi/2}\,,\quad\ 
S_{AC} = S_{\pi/4} + S_{\pi/4}\,,\quad\  S_{ABC} = S_{\pi/4}\,, && 
\eea
with $S_A\equiv S[A]$ etc.
Using these, we first consider tripartite information which
evaluates to
\bea\label{tripartiteI3}
I_3^t[A,B,C] &=& S_A + S_B + S_C - S_{AB} - S_{BC} - S_{AC} + S_{ABC} \nn\\
&=& 2S_{\pi/4} - 2S_{\pi/2} = i{l\over 2G_3}\log 2 \qquad \Rightarrow\quad
{\rm Im}\, I_3^t[A,B,C] > 0\,.
\eea
Note that the real parts as well as the imaginary divergent terms
cancel completely between the various terms, leaving behind $I_3^t$
above. This appears unrecognizable in de Sitter per se, but in fact
it is consistent via analytic continuation with tripartite information
in $AdS$ being negative. Indeed, noting $il\ra -L$, we see that
$I_3^t$ above in $dS$ continues to the negative definite value
$I_3^{AdS}\sim -{L\over G_3}<0$, consistent with \cite{Hayden:2011ag}.
Similar observations can be made for mutual time-information (\ref{MI-t})
and its $AdS$ analytic continuation, where we expect positivity
(as can be seen from the leading scaling ${L\over G_3}$ of the first
term in (\ref{MI-t}) upon continuation).

Likewise the strong subadditivity relations, again recalling
\cite{Hayden:2011ag}, become
\bea\label{SSB12dS}
&& SSB_1^t = S_{AB} + S_{BC} - S_{ABC} - S_B = -i{l\over 2G_3}\log 2\,,\nn\\
&& SSB_2^t = S_{AB} + S_{BC} - S_{A} - S_C = -i{l\over 2G_3}\log 2\,,
\eea
using (\ref{SABC-pi/4}). This again is
consistent with the known (real) positivity of these inequalities
in $AdS$, as can be seen from the analytic continuation $il\ra -L$
and the comments above.
As another example, consider the three disjoint boundary subregions
$A, B, C$, in Figure~\ref{fig3} (dual to the green, violet, blue
bulk subregions), which together make up the full circle
($\theta_\infty={\pi\over 3}$ each): then\ $S_A=S_B=S_C=S_{\pi/3}$,\
$S_{ABC}=0$ and $S_{AB}=S_C$ (since $AB$ is
complementary to $C$) etc, so that $I_3^t=0$. The strong subadditivity
relations give\ $SSB_1^t=S_{\pi/3}$ and $SSB_2^t=0$. These can also be
seen to be consistent with the $AdS$ entropy positivity relations
under the analytic continuation. These thus suggest\
${\rm Re}\,SSB_{1,2}^t\geq 0,\ \ {\rm Im}\,SSB_{1,2}^t\leq 0$.

Finally, consider four disjoint, adjacent subregions $A, B, C, D$,
each with $\theta_\infty={\pi\over 4}$: each has spread
$2\theta_\infty={\pi\over 2}$ so together they make up the full
circle. Then clearly
\be
S[A]+S[B]=S[C]+S[D]\,;\quad S[A\cup B]=S[C\cup D]=S^{IR}\,;\qquad\quad
\ee
$A\cup C$ has inner and outer extremal surfaces defined by
the complement subregions $B, D$, so 
\be
I_t[A,C] = S[A]+S[C]-S[A\cup C] = S[A]+S[C]-S[B]-S[D] = 0\,.
\ee
So antipodal dijsoint subregions such as $A, C$ have vanishing mutual
time-information (in contrast with (\ref{MI-t}) for adjacent
subregions). Note that the top timelike parts of the $A, C$ extremal
surfaces are disjoint (meeting at the $r=l$ plane): however the bulk
subregions within the hemisphere have overlap so are not disjoint.

Thus the $dS_3$ analogs of the entropy relations while being
complex-valued and apparently novel in fact intricately encode the
known entropy relations satisfied by the $AdS$ RT/HRT areas. From
the structure of the areas (\ref{IRsurfAdSdS}), (\ref{IRsurfAdSdS4}),
we expect that these sorts of inequalities will hold in higher
dimensional $dS$ as well. It is of course fascinating to ask what
these complex-valued entropy inequalities, \ie\ \
${\rm Re}\, I_t[A,B] \geq 0,\ \ {\rm Im}\, I_t[A,B]\leq 0, \ \
{\rm Im}\, I_3^t[A,B,C] \geq 0$,\ \ etc mean intrinsically in de
Sitter space.

\subsubsection{Qubit systems and pseudo-entropy/time-entanglement relations}

To put in perspective the above de Sitter extremal surface
area-entropy relations and inequalities, let us consider
time-entanglement/pseudo-entropy in simple quantum mechanical qubit
systems, in part recalling the discussions in \cite{Narayan:2023ebn}.

Consider a thermofield-double-type
initial state $|I\ran=\sum_{i=1,2}c_{ii}|ii\ran$ and its time-evolved final
state $|F\ran=\sum_ic_{ii} e^{-iE_{ii}t}|ii\ran$, where the Hamiltonian is
$H=E_{ij}|ij\ran\lan ij|$ with energy eigenvalues $E_{ij}$ for basis
states $|ij\ran$ (and we assume $E_{12}=E_{21}$). Then towards
evaluating pseudo-entropy, we note the normalized transition matrix and
its partial traces:
\bea
&& \rho_t={|F\ran\lan I|\over {\rm Tr} (|F\ran\lan I|)} 
= {1\over\sum_j |c_{jj}|^2e^{-iE_{jj}t}}\Big( |c_{11}|^2e^{-iE_{11}t}|11\ran\lan 11|
+ c_{11} c_{22}^*e^{-iE_{11}t}|11\ran\lan 22| \nn\\
&& \qquad\qquad\qquad\qquad\qquad\qquad\qquad\quad
+\, c_{11}^* c_{22}e^{-iE_{22}t}|22\ran\lan 11|
+ |c_{22}|^2e^{-iE_{22}t}|22\ran\lan 22| \Big)\,,\qquad\ \\
&& \rho_t^1 = {\rm Tr}_2\,\rho_t = {1\over\sum_j |c_{jj}|^2e^{-iE_{jj}t}}
\Big( |c_{11}|^2e^{-iE_{11}t}|1\ran\lan 1| +
c_{22}|^2e^{-iE_{22}t}|2\ran\lan 2| \Big)
= \rho_t^2 = {\rm Tr}_1\,\rho_t\,.\ \ \nn
\eea
$\rho_t$ can be equivalently regarded
\cite{Narayan:2022afv,Narayan:2023ebn} as the time evolution operator
with projection onto the initial state $|I\ran$\ (and $\rho_t^1,\ \rho_t^2$,
partial traces thereof),
\be
\rho_t = {{\cal U}(t)|I\ran\lan I|\over{\rm Tr} ({\cal U}(t)|I\ran\lan I|)}\,,
\qquad {\cal U}(t) = e^{-iHt}\,,\qquad |F\ran = {\cal U}(t)|I\ran\,,
\ee
Using the normalization $|c_{11}|^2+|c_{22}|^2=1$ and redefining\
$|c_{11}|^2=x$\ recasts this as
\bea\label{rhot12St12}
&& \rho_t^2 = \rho_t^1 = {1\over x+(1-x)e^{i\theta}}
\Big(x|1\ran\lan 1|+ (1-x)e^{i\theta}|2\ran\lan 2| \Big)\,,\qquad
\theta=-(E_{22}-E_{11})t=-\Delta E\,t\,,\nn\\
&& S_t^2 = S_t^1 = -{x\over x+(1-x)e^{i\theta}}\log {x\over x+(1-x)e^{i\theta}}
- {(1-x)e^{i\theta}\over x+(1-x)e^{i\theta}}
\log {(1-x)e^{i\theta}\over x+(1-x)e^{i\theta}}\,,\qquad
\eea
with $S_t^2=S_t^1$ the pseudo-entropy of the reduced transition matrix
corresponding to states $|I\ran, |F\ran$.
In the vicinity of $t=0$, we obtain
\bea\label{rhot12St12:t=0}
&& \qquad\qquad\qquad\qquad\qquad
S_t^1(t) \sim S_t^1(0) + {d\over dt}S_t^1(0)\,t \equiv S_0\,,\nn\\
&& S_t^1(0) = -x\log x - (1-x)\log (1-x)\,,\qquad
{d\over dt}S_t^1(0) = -i\,\Delta E\, x(1-x)\log {x\over 1-x}\,.\qquad
\eea
We see that $S_t^1(0)$ is the entanglement entropy of the initial state
at $t=0$ and is positive definite; however its (pure imaginary) time
derivative ${d\over dt}S_t^1(0)$ switches sign depending on whether
$x>{1\over 2}$ or $x<{1\over 2}$\,. Thus $S_0$ satisfies
${\rm Re} S_0>0$ and ${\rm Im}S_0 \gtrless 0$ for generic
$x\neq {1\over 2}$\,.

For this 2-qubit system with transition matrix $\rho_t$, let us
calculate mutual time-information (contrast (\ref{MI-t}). With
$\rho_t$ regarded as a $2\times 2$-matrix, we have ${\rm det}\,\rho_t=0$
so the eigenvalues are $0, 1$, and the associated entropy vanishes,
\ie\ $S_t=0$. So in the vicinity of $t=0$, we obtain
\be
I_t[1,2] = S_t^1 + S_t^2 - S_t \sim 2S_0\quad\Rightarrow\quad
{\rm Re} I_t>0\,,\quad {\rm Im} I_t \gtrless 0\,.
\ee
Thus the imaginary part of mutual time-information has no definite
sign for $x\neq {1\over 2}$\ (with fixed $\Delta E$). Evaluating $I_t$
near $t=0$ illustrates this sharply and explicitly, but it can
also be evaluated numerically for general $\theta$ (\ie\ $t$) using
(\ref{rhot12St12}) verifying the sign non-definiteness.
The maximally entangled case $x={1\over 2}$ has\ \
$S_t^2=S_t^1=-{1\over 1+e^{i\theta}}\log {1\over 1+e^{i\theta}}
- {1\over 1+e^{-i\theta}}\log {1\over 1+e^{-i\theta}}\equiv \al+\al^*$\
which is real-valued and positive for all $\theta$ (\ie\ $t$)\
\cite{Narayan:2022afv,Narayan:2023ebn}.

Now consider a 3-qubit system and a GHZ-type initial state $|I\ran$
and its time-evolved final state $|F\ran$, with Hamiltonian
$H=\sum E_{ijk}|ijk\ran$,\ and normalization $|c_{111}|^2+|c_{222}|^2=1$,
\be
|I\ran = c_{111}|111\ran+c_{222}|222\ran\,,\qquad
|F\ran = c_{111}e^{-iE_{111}t}|111\ran+c_{222}e^{-iE_{222}t}|222\ran\,,
\ee
Defining $|c_{111}|^2\equiv x$,\ the normalized transition matrix
and its various partial traces are
\bea
&&\!\!\!\! \rho_t^{123}={|F\ran\lan I|\over {\rm Tr} (|F\ran\lan I|)}\,,
\qquad\qquad \theta=-(E_{22}-E_{11})t=-\Delta E\,t\,, \nn\\
&&\!\!\!\! \rho_t^1={\rm Tr}_{23}\,\rho_t^{123}={1\over x+(1-x)e^{i\theta}}
\Big(x|1\ran\lan 1|+ (1-x)e^{i\theta}|2\ran\lan 2| \Big)\,,\qquad
\rho_t^2=\rho_t^3=\rho_t^1\,, \\
&&\!\!\!\! \rho_t^{12}={\rm Tr}_{3}\,\rho_t^{123}
={1\over x+(1-x)e^{i\theta}}
\Big(x|11\ran\lan 11|+ (1-x)e^{i\theta}|22\ran\lan 22| \Big)\,,\qquad
\rho_t^{23}=\rho_t^{13}=\rho_t^{12}\,.\nn
\eea
Operationally it is clear that each of the reduced transition matrices
are of the form for the 2-qubit case (\ref{rhot12St12}), so their
pseudo-entropies are all of the form $S_t^1$ there. It can be seen
that ${\rm det}\,\rho_t^{123}=0$ and 
the associated pseudo-entropy $S_t^{123}$ vanishes (the only eigenvector
is $|I\ran$). Thus finally we
evaluate\
$I_3^t[1,2,3] = S_t^1+S_t^2+S_t^3-S_t^{23}-S_t^{13}-S_t^{12}+S_t^{123}=0$\
so the tripartite information for this 3-qubit system vanishes.
We can also likewise evaluate the strong subadditivity relations here,
and contrast with (\ref{SSB12dS}): this gives
\be
SSB_1^t = S_t^{12} + S_t^{23} - S_t^{123} - S_t^2 = S_t^1\,,
\ee
while $SSB_2^t = S_t^{12} + S_t^{23} - S_t^{1} - S_t^3 = 0$.
Analysing in the vicinity of $t=0$ using (\ref{rhot12St12:t=0}) shows
\be
{\rm Re}\,SSB_1^t>0\,,\qquad {\rm Im}\,SSB_2^t \gtrless 0\,;
\qquad\quad x\neq {1\over 2}\,,
\ee
so the imaginary part of $SSB_1^t$ switches sign depending on
$x\gtrless {1\over 2}$ (for fixed $\Delta E$).

One can also analyse the entropy structures of the time evolution
operator as another diagnostic of entanglement-like structures with
timelike separations, as in \cite{Narayan:2022afv,Narayan:2023ebn}.
For the 2-qubit case the normalized time evolution operator is
\be
\rho_t = \sum_{i,j}
          {e^{-iE_{ij}t}\over \sum_{kl} e^{-iE_{kl}t}}\, |ij\ran\lan ij| 
\,=\, {|11\ran\lan 11|
+ e^{i\theta_1}|22\ran\lan 22| + e^{i\theta_2} ( |12\ran\lan 12|
+ |21\ran\lan 21| ) \over 1+e^{i\theta_1}+2e^{i\theta_2}}
\ee
with two arbitrary phases comprising
$\theta_1\equiv -(E_{22}-E_{11})t\,,\ \theta_2\equiv -(E_{12}-E_{11})t$.
Taking partial traces gives the reduced time evolution operator for
the remaining index,
\be
\rho_t^2 = \rho_t^1 =
    {1\over 1+e^{i\theta_1}+2e^{i\theta_2}} \Big( \big(1+e^{i\theta_2}\big)
|1\ran\lan 1| + \big(e^{i\theta_1}+e^{i\theta_2}\big) |2\ran\lan 2| \Big)\,.
\ee
The associated von Neumann entropies for $ \rho_t^1, \rho_t^2$ and
$\rho_t$ are \cite{Narayan:2023ebn}
\bea
&& S_t^{1,2} = -{1+e^{i\theta_2}\over 1+e^{i\theta_1}+2e^{i\theta_2}}\,
\log {1+e^{i\theta_2}\over 1+e^{i\theta_1}+2e^{i\theta_2}}\, -\,
{e^{i\theta_1}+e^{i\theta_2}\over 1+e^{i\theta_1}+2e^{i\theta_2}}\,
\log {e^{i\theta_1}+e^{i\theta_2}\over 1+e^{i\theta_1}+2e^{i\theta_2}}\ , \nn\\
&& S_t = -{1\over 1+e^{i\theta_1}+2e^{i\theta_2}}\,
\log {1\over 1+e^{i\theta_1}+2e^{i\theta_2}}\,
-\,{e^{i\theta_1}\over 1+e^{i\theta_1}+2e^{i\theta_2}}\,
\log {e^{i\theta_1}\over 1+e^{i\theta_1}+2e^{i\theta_2}}\, \qquad \nn\\
&& \qquad\qquad\qquad\ -\, {2e^{i\theta_2}\over 1+e^{i\theta_1}+2e^{i\theta_2}}\,
\log {e^{i\theta_2}\over 1+e^{i\theta_1}+2e^{i\theta_2}}\,, 
\eea
and mutual time-information becomes\
$I_t[1,2] = S_t^1+S_t^2-S_t = 2S_t^1-S_t$ which is complex-valued in
general (apart from specific special cases).
By numerically evaluating this for generic $\theta_{1,2}$, it can be
seen that ${\rm Re}\,I_t<0$ while the imaginary part exhibits both
signs as $t$ varies (for fixed $E_{ij}$). Evaluating near $t=0$ shows\
$I_t[1,2](0)=0$,\ ${d\over dt}I_t[1,2](0)=0$,\ with
the leading nonzero terms being\
${\rm Re}\,I_t[1,2]= -{1\over 32}(E_{11}+E_{22}-2E_{12})^2t^2+\ldots$ and
$i{\rm Im}\,I_t[1,2]=i{1\over 1536}(E_{22}-E_{11})^2(E_{11}+E_{22}-2E_{12})t^5+\ldots$,\ verifying the sign-nondefiniteness of the imaginary part.

The above qubit examples are quite simple in structure but serve to
explicitly illustrate the behaviour of
pseudo-entropy/time-entanglement relations and inequalities. More
complicated qubit examples quickly become very intricate in their
entropy structures.  It is clear that these simple qubit examples in
quantum mechanics are quite different in their pseudo-entropy
inequalities behaviour compared with the properties we saw for de
Sitter extremal surfaces earlier. In some ways this is not surprising:
it reflects the striking differences found in \cite{Hayden:2011ag}
with regard to entropy inequalities for the $AdS$ RT surface areas and
entanglement entropies in theories with gravity duals, in
comparison with qubit systems and generic CFTs. The (generically)
complex-valued nature of pseudo-entropies makes the entropy relations
and inequalities more intricate overall: it would be interesting to
understand the underlying organizational principles here, especially
in nonunitary theories with $dS$-like gravity duals.

\subsection{Antipodal observers at $I^+$ to codim-2 surfaces}\label{sec:dSobs}

So far we have been discussing codim-2 extremal surfaces, which appear
natural from the point of view of Euclidean time slices of the dual
boundary theory. From the bulk point of view, one might imagine asking
for natural subregions at $I^+$ being codim-0, \ie\ smaller regions of
the $S^d$ at $I^+$ (in $dS_{d+1}$). For instance, for $dS_3$ with the
$S^2$ at $I^+$, this would lead to cap-like subregions on the $S^2$
which lead to maximal subregions being hemispheres. These would lead
to codim-1 surfaces anchored at the boundary of these subregions.

\begin{figure}[h] 
\hspace{3.5pc}
\includegraphics[width=6pc]{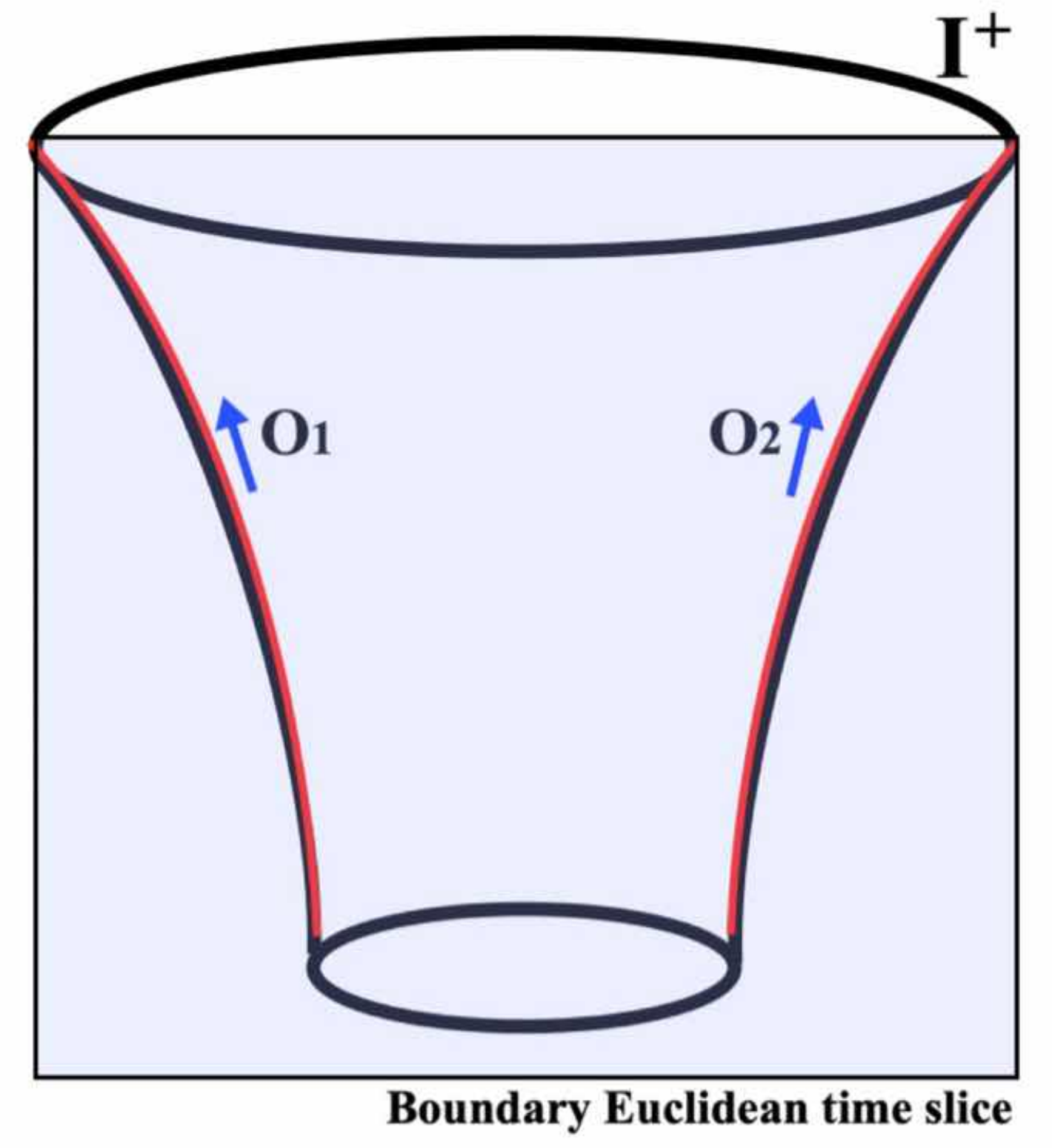}
\hspace{2pc}
\begin{minipage}[b]{23pc}
\caption{{ \label{fig0}
    \footnotesize{
      Antipodal observers $O_1$ and $O_2$ at $I^+$ in $dS$,
      and the boundary Euclidean time slice they define. This gives
      rise to maximal subregions with corresponding IR extremal
      surfaces.\newline
        }}}
\end{minipage}
\end{figure}

Instead consider two antipodal observers at $I^+$, not in causal
contact, and unable to communicate with each other.  For simplicity
consider them to be stationary, so they are represented by timelike
trajectories parametrized by global $dS$ time $\tau$ at the North and
South Poles, with all other angular coordinates fixed.  These
antipodal observers are ``maximally separated'', whereas two generic
observers are not.  While either observer (at either North or South
Pole) can only access information pertaining to half the space, the
two antipodal observers together provide a complete description of the
entire de Sitter space (these are best regarded as metaobservers in
the sense of \cite{Witten:2001kn}).

These two antipodal points define a plane normal to $I^+$: this is a
natural ``vertical'' bulk slice which is essentially a boundary
Euclidean time slice corresponding to an equatorial plane of the
$S^{d-1}$. Any two antipodal observers define such equatorial boundary
Euclidean time slices which are all equivalent: so there is a
continuum $S^{d-1}$ of such antipodal observers.\ Note also that from
(\ref{dSglobaleqSd-dSstattconst}), the $t=const$ slice in the $dS$
static coordinatization defines the same boundary Euclidean time slice
as a generic equatorial plane in global $dS$ obtained from two
antipodal observers.

Now these two observers will draw RT/HRT surfaces going into the time
direction (no single local patch suffices): their areas are perhaps
best regarded as meta-observables as described in
\cite{Witten:2001kn}. These now are future-past or no-boundary
extremal surfaces, with all their properties: in particular the
no-boundary extremal surface stretching out from one observer only
returns to the other observer (after turning around in the bottom
Euclidean hemisphere). The maximal subregions here are $S^{d-1}$
hemispheres so the extremal surface is anchored on the $S^{d-2}$
boundary and dips into time, \ie\ wrapping $S^{d-2}\times \tau$ which
gives a codim-2 surface.

In the static coordinatization with boundary topology $R_t\times
S^{d-1}$, the analytic continuation lands on the $t=const$ slice as we
have seen, from (\ref{nbdSstattoAdS2}): this defines ``preferred''
observers whose extremal surfaces wrap the $S^{d-2}$ boundary of the
$S^{d-1}$ subregions. For maximal subregions ($S^{d-1}$ hemispheres)
the IR extremal surfaces wrap $S^{d-2}\times\tau$ and are maximal:
calculationally these are identical to those in the equatorial plane
global slicing above.

It would be instructive to develop the point of view of these
antipodal (meta)observers further, possibly connecting the description
here of codim-2 extremal surfaces (which are akin to metaobservables
at the future boundary) with \cite{Chandrasekaran:2022cip}, as well as
\cite{Cotler:2023xku} more directly.

\section{No-boundary $dS$:\ Lewkowycz-Maldacena heuristically
  to pseudo-entropy}\label{sec:dSrepLM}

In this section, we will discuss how the analytic continuation from
$AdS$ to $dS$ maps the Lewkowycz-Maldacena replica argument
\cite{Lewkowycz:2013nqa} deriving the $AdS$ RT surfaces to a
corresponding heuristic version of the replica argument in $dS$.  Our
discussion is somewhat formal based mainly on the analytic
continuation. While a Lewkowycz-Maldacena argument for pseudo-entropy
appears in \cite{Nakata:2020luh} in the $AdS$ context, the
arguments here are specific to de Sitter which has several new
features.  Most notably the $dS/CFT$ dictionary $Z_{CFT}=\Psi_{dS}$
\cite{Maldacena:2002vr} implies that a replica on the boundary
partition function (for boundary entanglement entropy) amounts to a
replica on the bulk $dS$ Wavefunction (a single ket, rather than a
density matrix), so it is best interpreted as giving rise to bulk
pseudo-entropy, which is the main takeaway. Quantitatively, this
recovers the IR no-boundary $dS$ areas in (\ref{dS43rev}) from the
areas of appropriate codim-2 cosmic branes, as we will see.


First we note that under the analytic continuation from global $AdS$
to $dS$, the bulk $AdS$ partition function continues to the bulk $dS$
Wavefunction of the Universe $\Psi_{dS}$\ see \eg\
\cite{Maldacena:2002vr}, \cite{Harlow:2011ke},
\cite{Maldacena:2019cbz}). This Hartle-Hawking Wavefunction
\cite{Hartle:1983ai}, semiclassically, can be represented by a
gravitational path integral
\be
\Psi_{dS}[h_{ij},\phi_0] = \int_{nb}^{I^+} Dg\,D\phi\ e^{iS[g,\phi]}
\ee
obtained by summing over all $(d+1)$-manifolds with $h_{ij}$ as the
metric on the spatial slice at the future boundary $I^+$ (and
$\phi_0$ are final source boundary conditions for matter fields
$\phi$). There is
no time that appears in this representation of $\Psi_{dS}$: time 
implicitly enters in the sum over bulk spacetime histories. This
Wavefunction $\Psi_{dS}[h_{ij}]$ can be regarded as the amplitude
for creating the universe $M[h_{ij}]$ with these final boundary conditions
from ``nothing'', \ie\ satisfying the Hartle-Hawking no-boundary
condition (which is also imposed on the matter fields $\phi$).
Then it is reasonable to regard this Wavefunction $\Psi[h_{ij}]$
as the transition amplitude $\rho_t$ for\ ``nothing''$\ra$\ $M[h_{ij}]$.
No-boundary de Sitter space arises, as we have seen in \eg\
(\ref{nbdSstattoAdS}), by gluing the top Lorentzian part to the
bottom Euclidean hemisphere at the $r=l$ midplane.
Semiclassically, the Wavefunction is given by the action as\
\be
\Psi_{dS}\ \sim\ e^{iS^{(r>l)}}\,e^{S_E^{(r<l)}}
\ee
with the top Lorentzian part (with real $S^{(r>l)}$) being a pure phase 
while the bottom Euclidean hemisphere has real action.
Global $dS$ in (\ref{EAdSglobaldSglobal}) continues to the bottom
Euclidean hemisphere as\
$ds^2 = l^2 d\tau_E^2 + l^2 \cos^2\tau_E\,d\Omega_d^2$ with
$\tau=i\tau_E$,\ $0\leq\tau_E\leq {\pi\over 2}$\,: then the
Lorentzian $iS_{cl}$ continues to the Euclidean gravity action\ \
$S_E^{(r<l)}=-\int_{_{_{nbp}}}\sqrt{g}\,(R-2\Lambda)$\ \ in the hemisphere,
giving\ \
${1\over 2}\,{l^4V_{S^4}\over 16\pi G_4}\,{6\over l^2}={\pi l^2\over 2G_4}$\
\ for $dS_4$\ (the no-boundary point is $\tau_E={\pi\over 2}$ here).\ \
So the
probability for creating $dS$ from no-boundary nothingness is the
Hartle-Hawking factor\ $|\Psi_{dS}|^2=e^{dS.entropy}$\ from
$S_E^{(r<l)}$ \ \cite{Hartle:1983ai} (see also
\eg\ \cite{Bousso:1995cc,Bousso:1996au}, as well as
\cite{Maldacena:2011mk}). Further, see \cite{Maldacena:2019cbz} for
$dS_2$,\ \cite{Hikida:2022ltr} for the $dS_3$
case,\ \cite{Doi:2022iyj,Narayan:2022afv} for related discussions in
the context of the no-boundary extremal surface areas, and also the
recent \cite{Chakraborty:2023yed} in the context of the Wheeler-de
Witt equation.

We are attempting to closely mimic some of the $AdS$ arguments in
\cite{Dong:2016hjy} (see also \cite{Dong:2016fnf},
\cite{Dong:2013qoa}) extending
\cite{Lewkowycz:2013nqa}, as reviewed in \cite{Rangamani:2016dms} and
\cite{Callebaut:2023fnf} in extending to the $dS$ case here.  Since
the analytic continuation from $AdS$ to $dS$ maps
$Z_{bulk}\ra\Psi_{dS}$, the replica construction of
Lewkowycz-Maldacena \cite{Lewkowycz:2013nqa} in $AdS$ is now a
formulation on the Wavefunction, restricted to the boundary Euclidean
time slice (mapped from the constant time slice in $AdS$ where the
subregion is defined). The associated entropy is then the entropy of
the corresponding transition matrix, \ie\ the time-entanglement or
pseudo-entropy.

Let us first do a quick review of the $AdS$ case:  here,
we are considering a boundary subregion on a constant time slice. The
subregion boundary is then codim-2 so the bulk extremal surface
anchored on it is also codim-2 in the bulk. The replica space is an
$n$-fold branched cover over the boundary, branched over the subregion
boundary on the constant time slice (which thus makes the subregion
boundary codim-2).  The $\BZ_n$ replica symmetry permutes the $n$
boundary copies. The boundary replica space $M_n$ extends into a smooth
bulk replica space ${\cal B}_n$ which is to be regarded as a smooth
covering space wth replica boundary conditions for gluing the $n$
copies (cyclically). If we now consider the quotient
${\tilde{\cal B}_n}$ of the bulk space by the $\BZ_n$ replica symmetry
(so its boundary is $\del{\tilde{\cal B}_n}=M_n/\BZ_n=M_1$, the
original boundary space), then we encounter conical (orbifold)
singularities corresponding to $\BZ_n$ fixed points in the bulk
for $n\neq 1$. These curvature singularities are a reflection of the
fact that
the bulk quotient space is a solution to the bulk Einstein equations
only in the presence of a source with nontrivial backreaction.
Since the fixed points extend out from the subregion boundary, the
required source in question is a codim-2 (cosmic) brane with tension
${n-1\over n}{1\over 4G}$\ (giving deficit angle $2\pi-{2\pi\over n}$).
The replica quotient space ${\tilde{\cal B}_n}$ comprises $n$ copies
which are locally the same as a single copy, except for the fixed
point singularity: this is smoothed out by the codim-2 cosmic
(spacelike) brane source which has area $A$ (wrapping all the
transverse directions). Thus the smooth action $I_n$ can be written
as
\be
I_n = n I_1 + I_{brane} = n I_1 + {n-1\over n} {A\over 4G}
\ \ \xrightarrow{\, n=1+\epsilon\, }\ \ (1+\epsilon) I_1
+ \epsilon\,{A\over 4G}\,,
\ee
where we have written $n=1+\epsilon$ near $n=1$.
In the limit when $n\ra 1$, the cosmic brane becomes a tensionless
probe localized on the bulk RT/HRT surface.
Defining the bulk partition function $Z_n\equiv Z[{\tilde{\cal B}_n}]$
on the replica quotient space ${\tilde{\cal B}_n}$, the entropy via
replica is obtained as
\be
S = -\lim_{n\ra 1} n\del_n (\log Z_n - n\log Z_1)
= \lim_{n\ra 1} (1-n\del_n) \log Z_n = -\lim_{n\ra 1} (1-n\del_n) I_n
\ee
in the semiclassical approximation where\ $Z_n\sim e^{-I_n}$, with
$I_n$ the action. Also, $Z_n={\rm tr}\,\rho^n$, and $Z_1^n$ ensures
that the entropy pertains to the correct normalization\
${\rm tr}\,\rho=1$.
Thus as $n\ra 1$ (\ie\ $\epsilon\ra 0$), we obtain\ $S={A\over 4G}$
which is the area of the extremal RT/HRT entangling surface.

We can obtain a heuristic understanding of the replica formulation of
the entropy of the Wavefunction by following through (for simplicity)
the analytic continuation (\ref{IRsurfAdSdS}) for the IR extremal
surface in $dS$ on the $t=const$ slice, \ie\ for the maximal subregion
at $I^+$ (with area (\ref{IRsurfAdSdS4}) for $dS_4$ and
(\ref{IRsurfAdSdS3}) for $dS_3$). In this case, we are inserting twist
operators at the $\theta=\pm{\pi\over 2}$ endpoints of the
(hemispherical) maximal subregion at $I^+$ on the $t=const$ slice
(Figure~\ref{fig1}) and then constructing $n$ copies of the
Wavefunction $\Psi$ appropriately gluing the $n$ copies cyclically
with replica boundary conditions.\ Under the analytic continuation we
have
\be
Z_{bulk} \sim e^{-I_{bulk}}\ \ \longrightarrow\ \ \Psi_{dS}\sim e^{iS_{cl}}
\ee
for a single copy. In the replica case with $n$ copies, we expect
\be
   {Z_n\over Z_1^n}\ \ \longrightarrow\ \ {\Psi_n\over \Psi_1^n}\,.
\ee
$\Psi_n$ is the Wavefunction on the quotient bulk replica space with
nontrivial replica boundary conditions on the future boundary (and
$\Psi_1$ for a single copy).\  Semiclassically, we have
\be
Z_n\sim e^{-I_n}\longrightarrow \Psi_n\sim e^{iS_n}\,;\qquad\quad
-I_n\ \ra\ iS_n = iS_n^{(r>l)} + S_E^{(r<l)}\,,
\ee
where, under (\ref{nbdSstattoAdS}), the semiclassical
$AdS$ action maps to Lorentzian $dS$ in the top Lorentzian $r>l$ part
(with pure imaginary $iS_n$) and the bottom Euclidean hemisphere
for $r<l$ (with real $S_E$). The codim-2 brane source for $n\neq 1$
has nontrivial (part Euclidean, part Lorentzian) time evolution:
in the $n\ra 1$ limit it satisfies the no-boundary condition and
wraps the will-be no-boundary $dS$ extremal surface. Now
continuing from above gives $-I_{brane}\ra {1-n\over n} {A_{brane}\over 4G}$
so we obtain the area in $dS$ (with $n\ra 1$) as
\be\label{LMbraneAreadS}
S_t = \lim_{n\ra 1} (1-n\del_n) \log \Psi_n
= \lim_{n\ra 1} (1-n\del_n)\, {1-n\over n}\, {A_{brane}\over 4G}
= {A^{dS}_{brane}\over 4G}\,.
\ee
This is not a spacelike brane here localized on a spacelike RT/HRT
surface but a time-evolving brane localized on the part Euclidean,
part Lorentzian no-boundary $dS$ extremal surface. The area is complex
(with (\ref{IRsurfAdSdS4}), (\ref{IRsurfAdSdS3}), for $dS_4$ and
$dS_3$ respectively). This is a replica formulation of the entropy of
the Wavefunction (restricted to this boundary Euclidean slice): this
is not hermitian, unlike an ordinary density matrix. It is thus best
interpreted as time-entanglement or pseudo-entropy, rather than
ordinary spatial entanglement entropy.

It is reasonable that this entropy is complex-valued, with the top
Lorentzian timelike part being pure imaginary in particular. In the
Lewkowycz-Maldacena replica formulation, the entropy arises as the
area of the codim-2 brane created from nothing. The amplitude for this
process would be divergent if the Lorentzian part (which goes all the
way to late times) were real: as it stands, the timelike parts give
pure phases which cancel in the probability. We obtain a finite value
since the real part of this entropy is bounded, arising as it does
from the brane wrapping the Euclidean hemisphere which has a maximal
size (set by de Sitter entropy).

The above arguments are of course consistent with the entropy here
being the analog of entanglement entropy in the nonunitary
Euclidean CFT dual to $dS$. Since $Z_{CFT}=\Psi_{dS}$ via $dS/CFT$,
this replica formulation on $Z_{CFT}$ amounts to an analytic
continuation of that on $Z_{CFT}=Z_{bulk}^{AdS}$. The analytic
continuations (\ref{IRsurfAdSdS4}), (\ref{IRsurfAdSdS3}), of the
extremal surface areas in $dS_4,\ dS_3$, respectively can be seen to be
the analytic continuations of the entanglement entropy for the
relevant maximal subregions in $AdS$ (with sphere boundary). This is
broadly consistent with the early discussions on holographic
entanglement in $dS/CFT$ by analytic continuations
\cite{Narayan:2015vda,Narayan:2015oka}, \cite{Sato:2015tta}, and
with appropriate analytic continuations of entanglement entropy
in 2-dim CFT \cite{Holzhey:1994we,Calabrese:2004eu,Calabrese:2009qy}.
More broadly, the entire construction of the codim-2 bulk extremal
surfaces
\cite{Narayan:2015vda,Narayan:2017xca,Narayan:2020nsc,Narayan:2022afv}
on boundary Euclidean time slices and subregions therein was designed
to mirror boundary entanglement entropy in the exotic Euclidean CFT
dual to $dS$. The corresponding boundary replica on $Z_{CFT}$ then
maps via the above heuristic arguments to a bulk replica on
$\Psi_{dS}$ resulting in the corresponding pseudo-entropy.

It is to be noted that the arguments above leading to
(\ref{LMbraneAreadS}) are obtained via analytic continuation from
$AdS$. As discussed above, these appear reasonable for the perspective
here, give the expected no-boundary $dS$ extremal surface areas and
are consistent with expectations of the boundary nonunitary CFT.
Assumptions implicit here are that $I_n$ is indeed the dominant saddle
when analysed directly in the de Sitter case and is well-behaved for
the purposes here: it would be instructive to analyse this more
explicitly.

In addition, it would be nice to understand if there are alternative
ways to formulate the Lewkowycz-Maldacena replica arguments, without
relying on the analytic continuation. It is then conceivable that one
ends up with not one copy of the Wavefunction but with both a ket and
a bra, appropriately glued in some appropriate Schwinger-Keldysh
replica path integral. It might then (naively) seem that such a
formulation would be analogous to a replica on $Z_{CFT}^*Z_{CFT}$ and
thus be manifestly positive with no timelike component, unlike the
above (see \cite{Narayan:2016xwq} for some related comments in the
context of ghost CFT replicas and $dS/CFT$).

\section{On future-past entangled states
  and time-evolution}\label{sec:fpTFD}

We now attempt to draw analogies with the arguments in
\cite{VanRaamsdonk:2010pw}, \cite{VanRaamsdonk:2009ar}, to understand
certain aspects of future-past entangled states and
connectedness of time evolution and thereby emergence of time.
We will mostly focus on ordinary quantum mechanics here.

It was argued in \cite{Narayan:2017xca,Narayan:2020nsc},
that future-past de Sitter extremal surfaces stretching between
$I^\pm$ suggest future-past entanglement between two copies of the
dual Euclidean CFT in thermofield-double type entangled states,
in analogy with the eternal $AdS$ black hole \cite{Maldacena:2001kr}.
Certain properties of such states were further studied in
\cite{Narayan:2022afv}. With this in mind, 
consider doubling the Hilbert space ${\cal H}$ of some quantum system,
described by Hamiltonian eigenstates $|I\ran$ with energies $E_I$. The
past copy, say at $t=0$, will be denoted by ${\cal H}_P$ and the future
copy at say $t=T$, is ${\cal H}_F$, and is essentially isomorphic to
${\cal H}_P$, with the isomorphism map being the time evolution
operator ${\cal U}(t)$. The isomorphism is characterized by the
standard time evolution of the Hamiltonian eigenstates, \ie\
\be
{\cal U}(t) = e^{-iHt}\,:\quad |I\ran_F = e^{-iE_IT}|I\ran_P
\ee
Then generic states in ${\cal H}^{(1)}\times {\cal H}^{(2)}
= {\cal H}_F\times {\cal H}_P$ are
\be\label{genstPsifp}
|\Psi_g\ran_{FP} = \sum_{I,J} c_{IJ} |I\ran_F^{(1)} |J\ran_P^{(2)}
= \sum_{I,J} c_{IJ} e^{-iE_IT} |I\ran_P^{(1)} |J\ran_P^{(2)}
\equiv |\Psi_g\ran_{PP}
\ee
These sorts of states involve timelike separations between the states
from the $F$ and $P$ copies, which translates to time evolution phases
when written in terms of corresponding states in
${\cal H}_P^{(1)}\times {\cal H}_P^{(2)}$.\
There are a few noteworthy points here, in considering such a
doubling of the Hilbert space.

Firstly, ${\cal H}_F\times {\cal H}_P$\ (motivated by future-past $dS$
surfaces is somewhat different qualitatively from factorizing a
Hilbert space as
${\cal H}={\cal H}_L\times {\cal H}_R$ (as in the $CFT_L\times CFT_R$
dual to the eternal $AdS$ black hole): in that case there is no
causal connection between the components ${\cal H}_L$ and ${\cal H}_R$,
which are independent. In the current case ${\cal H}_F$ is the
time evolution of ${\cal H}_P$, \ie\ the time evolution operator
acts as the isomorphism.
So a state $|\Psi\ran_{FP}$ is isomorphic to a corresponding state
$|\Psi\ran_{FF}$. Since the time evolution phases cancel, the state
$|\Psi\ran_{FP}$ is future-past entangled if the corresponding
state $|\Psi\ran_{FF}$ is entangled in the ordinary sense on the
(constant) late time slice $F$.

To elaborate, consider a factorized future-past state (\ref{genstPsifp})
and the partial trace over the past copy of the corresponding
future-past density matrix $\rho_{fp}$:
\be
|\Psi\ran_{fp} = |\psi_F\ran^{(1)} |\psi_P\ran^{(2)}
= {\cal U}(t)|\psi_P'\ran^{(1)} |\psi_P\ran^{(2)}
\quad\ra\quad
    {\rm Tr}_P\,\big(|\Psi\ran_{fp} \lan\Psi|_{fp}\big)
    = |\psi_F\ran^{(1)}\lan\psi_F|^{(1)}
\ee
which is pure at the future time slice, with zero entropy. Note also
that the time evolution operator ${\cal U}(t)$ has disappeared in the
future-past density matrix. This calculation thus does not care
whether $|\psi_F\ran$ and $|\psi_P\ran$ belong in the same Hilbert
space: the state effectively is $|\psi_F\ran^{(1)} |\psi_P\ran^{(2)}$
with two disconnected spaces $1, 2$.

On the other hand, a non-factorizable future-past state such as that in
(\ref{genstPsifp})
has generic nonzero coefficients $c_{IJ}$ giving the trace over the past
copy of the future-past density matrix
$\rho_{fp} = |\Psi\ran_{FP} \lan\Psi|_{FP}$ as
\bea\label{fp-DM}
{\rm Tr}_P\, \big(|\Psi\ran_{FP} \lan\Psi|_{FP}\big)
&=& {\rm Tr}_P\, \sum_{IJ,KL} c_{IJ} c^*_{KL} |I\ran_F^{(1)}
|J\ran_P^{(2)} \lan K|_F^{(1)} \lan L|_P^{(2)}
\nn\\
&=& \sum_{IJ,KL} \delta_{JL} c_{IJ} c^*_{KL} |I\ran_F^{(1)} \lan K|_F^{(1)}
= \sum_{I,K} c_{IJ} c^*_{KJ} e^{-iE_IT} e^{iE_KT} |I\ran_P^{(1)} \lan K|_P^{(1)}
\qquad
\eea
The form in terms of the $F$ copies alone is seen to be entirely
positive: this is a mixed state with nonzero (ordinary) entanglement
entropy at the future time-slice $T$ if the future-past state is
non-factorizable, \ie\ entangled.

The time evolution operator can be obtained as a reduced transition
matrix with thermofield double type states using basis states $|i\ran$
as below (with the normalization done later):
\bea\label{tE-TrTM}
&&\qquad  |\psi_I\ran = \sum_{i=1}^N |i\ran_P^{(1)} |i\ran_P^{(2)}\,, \qquad
|\psi_{FP}\ran = \sum_{i=1}^N |i\ran_F^{(1)} |i\ran_P^{(2)}
= \sum_{i=1}^N e^{-iE_it} |i\ran_P^{(1)} |i\ran_P^{(2)}\,, \nn\\
&& 
{\cal U}(t) = \sum_{i=1}^N e^{-iE_it} |i\ran_P\lan i|_P
= \sum_{i=1}^N |i\ran_F\lan i|_P \nn\\
&& \qquad\qquad\qquad\qquad
= {\rm Tr}_2\, \big( |\psi_{FP}\ran \lan \psi_I| \big)
= \sum_{i,j} \delta_{ij}^{(2)}
\big( |i\ran_F^{(1)}|i\ran_P^{(2)}\, \lan j|_P^{(1)}\lan j|_P^{(2)} \big)\,.
\eea
This is distinct in detail from the way the time evolution operator
was obtained by the doubled Hilbert space states in
\cite{Doi:2022iyj}, \cite{Narayan:2023ebn}. In particular here
$|\psi_{FP}\ran\in {\cal H}_F\times {\cal H}_P$ for nonzero $t$ so it
is a future-past entangled state.
The time evolution operator acts as the map from the past space to the
future one which is not independent from the past one. Thus the
existence of the time evolution operator appears to be tantamount to
the existence of future-past entangled states of the form
$|\psi_{FP}\ran$. To elaborate, the time evolution operator implies
connectedness of the future and past slices by time evolution: the
existence of future-past entangled states is thus equivalent to this
time-connectedness.

It is worth noting two aspects of the above (in part elaborating on
the corresponding discussions in \cite{Narayan:2022afv}). One is the
fact that the time evolution operator itself can be obtained by
partial trace over the second copy in a doubled Hilbert space
construction involving future-past entangled states $|\psi_{FP}\ran$
as in (\ref{tE-TrTM}).  This contains non-positive structures, in
particular the time evolution phases, and arises from just a single
copy of $|\psi_{FP}\ran$\ (in some sense, this dovetails with the
comments at the end of the previous section, sec.~\ref{sec:dSrepLM},
on Lewkowycz-Maldacena for a single copy of the $dS$ Wavefunction
giving rise to pseudo-entropy). The other pertains to constructing a
density matrix from two copies of $|\psi_{FP}\ran$, as in
(\ref{fp-DM}): this leads to entirely positive structures. It would be
interesting to understand the interplay between these two aspects more
elaborately.

In the bulk de Sitter space, the future-past surfaces connecting the
past and future time slices are essentially the extremal time
evolution trajectories of codim-2 branes: so their existence is
tantamount to time evolution connecting the future and past slices in
the same physical spacetime.  One might ask if these future-past
surfaces ``break'' and disconnect: it would appear that this would not
occur except if the spacetime itself ``tears''\ (for instance, a
sufficiently singular perturbation from $I^-$ might create a Big-Crunch
singularity and thereby destroy $I^+$, so timelike trajectories end at
the singularity). Our discussions here have been for small
fluctuations about de Sitter so the future-past surfaces appear
to always exist. As an aside, Schwarzschild de Sitter black holes,
regarded as other possible endpoints to strong perturbations, also
admit similar future-past surfaces stretching between $I^\pm$ and
dipping further in towards the black hole \cite{Fernandes:2019ige}
(but $dS/CFT$ interpretations are unclear for $SdS$).



\section{Discussion}\label{sec:Disc}

We have developed further the investigations in \cite{Narayan:2022afv}
on de Sitter space, extremal surfaces and time entanglement.  We
obtain the no-boundary de Sitter extremal surface areas as certain
analytic continuations from $AdS$ (in part overlapping with
\cite{Doi:2022iyj}) while also amounting to space-time rotations
geometrically. This is most easily seen (sec.~\ref{sec:dSacIR}) for
the IR extremal surface for maximal subregions in any $dS_{d+1}$\ and
more generally in greater detail for $dS_3$\ (sec.~\ref{sec:AdS3dS3}).
The structure of the extremal surfaces also suggests a geometric
picture of the time-entanglement or pseudo-entanglement wedge
(sec.~\ref{sec:subregDuality}).  In sec.~\ref{sec:dSobs}, we have
given certain simple arguments connecting two antipodal
(meta)observers and codim-2 surfaces. The analytic continuation
suggests a heuristic Lewkowycz-Maldacena formulation of the $dS$
extremal surface areas via analytic
continuation\ sec.~\ref{sec:dSrepLM}:\ in the bulk, this is now a
replica formulation on the Wavefunction (single copy) which suggests
interpretation as pseudo-entropy.  Finally we also discuss aspects of
future-past thermofield states and time evolution in quantum mechanics
in sec.~\ref{sec:fpTFD}, suggesting close relations between the time
evolution operator and future-past entangled states.  Several of the
discussions here are suggestive rather than conclusive: we hope to
understand them more firmly over time.

In sec.~\ref{sec:MITISSB}, we studied de Sitter analogs of some of the
well-known entropy relations satisfied by RT/HRT surfaces in $AdS$,
specifically mutual information, tripartite information and strong
subadditivity, focussing on the detailed area expressions in $dS_3$
for generic subregions. The complex-valued areas imply complex-valued
entropy inequalities, \ ${\rm Re}\, I_t[A,B] \geq 0,\ \ {\rm Im}\,
I_t[A,B]\leq 0,\ \ {\rm Im}\, I_3^t[A,B,C] \geq 0$\ etc, which appear
novel in themselves (and also show interesting differences compared
with pseudo-entropy in qubit systems). However as we saw via the $AdS$
analytic continuation, these in fact intricately encode the known
positivity properties in $AdS$ \cite{Hayden:2011ag}.  It is
fascinating to ask how the entropy relations are organized
intrinsically in $dS$: perhaps these encode interesting properties of
both de Sitter space and of pseudo-entropy for (nonunitary) theories
with $dS$-like holographic duals.

It is important to note that our discussions in
sec.~\ref{sec:subregDuality} for the time-entanglement or
pseudo-entanglement wedge in no-boundary de Sitter (resembling a
space-time rotation from the entanglement wedge in $AdS$) are simply
geometric, and based on the analytic continuation.  It would be
fascinating to ``derive'' these heuristic geometric observations more
directly from analogs in $dS/CFT$ of modular flow, relative entropy,
error correction codes and so on
\cite{Almheiri:2014lwa,Jafferis:2015del,Dong:2016eik}. It would appear
that this would amount to a detailed understanding of how entanglement
wedge reconstruction works in de Sitter. In this regard, we note that
the central dictionary $Z_{CFT}=\Psi_{dS}$ (naively) suggests the
bulk-boundary relations are perhaps more intricate here, with
nontrivial entry of $|\Psi_{dS}|^2$ and the area relations
(\ref{dS43rev2}) in the way bulk subregions are encoded from boundary
data at $I^+$. It would be interesting to develop our discussions here
incorporating these, and to possibly usefully adapt the discussions in
\cite{Miyaji:2021lcq} to the de Sitter case. We hope to address these
and related issues in the future.

\vspace{8mm}

{\footnotesize \noindent {\bf Acknowledgements:}\ \ It is a pleasure
  to thank Philip Argyres, Sumit Das, Daniel Harlow, Veronika Hubeny,
  Juan Maldacena, Alex Maloney, Shiraz Minwalla, Rob Myers, Suvrat
  Raju, Mukund Rangamani, Al Shapere, Andy Strominger and Mark van
  Raamsdonk for helpful discussions, and also Mukund Rangamani and
  Hitesh Saini for comments on a draft. I thank the Organizers of
  Strings 2023, Perimeter Institute, and the String Group, U.Kentucky,
  for hospitality while this work was in progress.  This work is
  partially supported by a grant to CMI from the Infosys Foundation.
}



\begin{thebibliography}{} 

\footnotesize{

\bibitem{Maldacena:1997re}
  J.~M.~Maldacena,
  ``The large N limit of superconformal field theories and supergravity,''
  Adv.\ Theor.\ Math.\ Phys.\  {\bf 2}, 231 (1998)
  [Int.\ J.\ Theor.\ Phys.\  {\bf 38}, 1113 (1999)]
  [arXiv:hep-th/9711200].

\bibitem{Gubser:1998bc}
  S.~S.~Gubser, I.~R.~Klebanov and A.~M.~Polyakov,
  ``Gauge theory correlators from non-critical string theory,''
  Phys.\ Lett.\  B {\bf 428}, 105 (1998)
  [arXiv:hep-th/9802109].

\bibitem{Witten:1998qj}
  E.~Witten,
  ``Anti-de Sitter space and holography,''
  Adv.\ Theor.\ Math.\ Phys.\  {\bf 2}, 253 (1998)
  [arXiv:hep-th/9802150].

\bibitem{Strominger:2001pn} 
  A.~Strominger,
  ``The dS / CFT correspondence,''
  JHEP {\bf 0110}, 034 (2001)
  [hep-th/0106113].

\bibitem{Witten:2001kn} 
  E.~Witten,
  ``Quantum gravity in de Sitter space,''
  [hep-th/0106109].

\bibitem{Maldacena:2002vr}
  J.~M.~Maldacena,
  ``Non-Gaussian features of primordial fluctuations in single field inflationary models,''
  JHEP {\bf 0305}, 013 (2003),\ 
  [astro-ph/0210603].

\bibitem{Anninos:2011ui} 
  D.~Anninos, T.~Hartman and A.~Strominger,
  ``Higher Spin Realization of the dS/CFT Correspondence,''
  Class.\ Quant.\ Grav.\  {\bf 34}, no. 1, 015009 (2017)
  doi:10.1088/1361-6382/34/1/015009
  [arXiv:1108.5735 [hep-th]].

\bibitem{Spradlin:2001pw} 
  M.~Spradlin, A.~Strominger and A.~Volovich,
  ``Les Houches lectures on de Sitter space,''
  hep-th/0110007.

\bibitem{Anninos:2012qw}
D.~Anninos,
``De Sitter Musings,''
Int. J. Mod. Phys. A \textbf{27}, 1230013 (2012)
doi:10.1142/S0217751X1230013X
[arXiv:1205.3855 [hep-th]].

\bibitem{Galante:2023uyf}
D.~A.~Galante,
``Modave lectures on de Sitter space \& holography,''
PoS \textbf{Modave2022}, 003 (2023)
doi:10.22323/1.435.0003
[arXiv:2306.10141 [hep-th]].

\bibitem{Gibbons:1977mu} 
  G.~W.~Gibbons and S.~W.~Hawking,
  ``Cosmological Event Horizons, Thermodynamics, and Particle Creation,''
  Phys.\ Rev.\ D {\bf 15}, 2738 (1977).
  doi:10.1103/PhysRevD.15.2738

\bibitem{Ryu:2006bv} 
  S.~Ryu and T.~Takayanagi,
  ``Holographic derivation of entanglement entropy from AdS/CFT,''
  Phys.\ Rev.\ Lett.\  {\bf 96}, 181602 (2006)
  [hep-th/0603001].

\bibitem{Ryu:2006ef} 
  S.~Ryu and T.~Takayanagi,
  ``Aspects of Holographic Entanglement Entropy,''
  JHEP {\bf 0608}, 045 (2006)
  [hep-th/0605073].

\bibitem{HRT} 
V.~E.~Hubeny, M.~Rangamani and T.~Takayanagi,
``A Covariant holographic entanglement entropy proposal,'' 
JHEP {\bf 0707} (2007) 062  [arXiv:0705.0016 [hep-th]].

\bibitem{Rangamani:2016dms}
M.~Rangamani and T.~Takayanagi,
``Holographic Entanglement Entropy,''
Lect. Notes Phys. \textbf{931}, pp.1-246 (2017)
Springer, 2017,
doi:10.1007/978-3-319-52573-0
[arXiv:1609.01287 [hep-th]].

\bibitem{Narayan:2015vda} 
  K.~Narayan,
  ``de Sitter extremal surfaces,''
  Phys.\ Rev.\ D {\bf 91}, no. 12, 126011 (2015)
  [arXiv:1501.03019 [hep-th]].

\bibitem{Narayan:2015oka} 
  K.~Narayan,
  ``de Sitter space and extremal surfaces for spheres,''
  Phys.\ Lett.\ B {\bf 753}, 308 (2016)
  [arXiv:1504.07430 [hep-th]].

\bibitem{Sato:2015tta} 
  Y.~Sato,
  ``Comments on Entanglement Entropy in the dS/CFT Correspondence,''
  Phys.\ Rev.\ D {\bf 91}, no. 8, 086009 (2015)
  [arXiv:1501.04903 [hep-th]].

\bibitem{Miyaji:2015yva} 
  M.~Miyaji and T.~Takayanagi,
  ``Surface/State Correspondence as a Generalized Holography,''
  PTEP {\bf 2015}, no. 7, 073B03 (2015)
  doi:10.1093/ptep/ptv089
  [arXiv:1503.03542 [hep-th]].

 \bibitem{Narayan:2017xca} 
 K.~Narayan,
 ``On extremal surfaces and de Sitter entropy,''
 Phys.\ Lett.\ B {\bf 779}, 214 (2018)
 [arXiv:1711.01107 [hep-th]].

\bibitem{Narayan:2020nsc}
K.~Narayan,
``de Sitter future-past extremal surfaces and the entanglement wedge,''
Phys. Rev. D \textbf{101}, no.8, 086014 (2020)
doi:10.1103/PhysRevD.101.086014
[arXiv:2002.11950 [hep-th]].

\bibitem{Doi:2022iyj}
K.~Doi, J.~Harper, A.~Mollabashi, T.~Takayanagi and Y.~Taki,
``Pseudoentropy in dS/CFT and Timelike Entanglement Entropy,''
Phys. Rev. Lett. \textbf{130}, no.3, 031601 (2023)
doi:10.1103/PhysRevLett.130.031601
[arXiv:2210.09457 [hep-th]].

\bibitem{Narayan:2022afv}
K.~Narayan,
``de Sitter space, extremal surfaces, and time entanglement,''
Phys. Rev. D \textbf{107}, no.12, 126004 (2023)
doi:10.1103/PhysRevD.107.126004
[arXiv:2210.12963 [hep-th]].

\bibitem{Hikida:2022ltr}
Y.~Hikida, T.~Nishioka, T.~Takayanagi and Y.~Taki,
``CFT duals of three-dimensional de Sitter gravity,''
JHEP \textbf{05}, 129 (2022)
doi:10.1007/JHEP05(2022)129
[arXiv:2203.02852 [hep-th]].

\bibitem{Hikida:2021ese}
Y.~Hikida, T.~Nishioka, T.~Takayanagi and Y.~Taki,
``Holography in de Sitter Space via Chern-Simons Gauge Theory,''
Phys. Rev. Lett. \textbf{129}, no.4, 041601 (2022)
[arXiv:2110.03197 [hep-th]].

\bibitem{Hartman:2013qma} 
  T.~Hartman and J.~Maldacena,
  ``Time Evolution of Entanglement Entropy from Black Hole Interiors,''
  JHEP {\bf 1305}, 014 (2013)
  [arXiv:1303.1080 [hep-th]].

\bibitem{Maldacena:2001kr} 
  J.~M.~Maldacena,
  ``Eternal black holes in anti-de Sitter,''
  JHEP {\bf 0304}, 021 (2003)
  [hep-th/0106112].

\bibitem{Arias:2019pzy} 
  C.~Arias, F.~Diaz and P.~Sundell,
  ``De Sitter Space and Entanglement,''
  Class.\ Quant.\ Grav.\  {\bf 37}, no. 1, 015009 (2020)
  doi:10.1088/1361-6382/ab5b78
  [arXiv:1901.04554 [hep-th]].

\bibitem{Arias:2019zug}
C.~Arias, F.~Diaz, R.~Olea and P.~Sundell,
``Liouville description of conical defects in dS$_4$, Gibbons-Hawking entropy as modular entropy, and dS$_3$ holography,''
JHEP \textbf{04}, 124 (2020)
doi:10.1007/JHEP04(2020)124
[arXiv:1906.05310 [hep-th]].

\bibitem{Cotler:2023xku}
J.~Cotler and A.~Strominger,
``Cosmic ER=EPR in dS/CFT,''
[arXiv:2302.00632 [hep-th]].

\bibitem{Cotler:2022weg}
J.~Cotler and A.~Strominger,
``The Universe as a Quantum Encoder,''
[arXiv:2201.11658 [hep-th]].

\bibitem{Nakata:2020luh}
Y.~Nakata, T.~Takayanagi, Y.~Taki, K.~Tamaoka and Z.~Wei,
``New holographic generalization of entanglement entropy,''
Phys. Rev. D \textbf{103}, no.2, 026005 (2021)
[arXiv:2005.13801 [hep-th]].

\bibitem{Mollabashi:2020yie}
A.~Mollabashi, N.~Shiba, T.~Takayanagi, K.~Tamaoka and Z.~Wei,
``Pseudo Entropy in Free Quantum Field Theories,''
Phys. Rev. Lett. \textbf{126}, no.8, 081601 (2021)
doi:10.1103/PhysRevLett.126.081601
[arXiv:2011.09648 [hep-th]].

\bibitem{Mollabashi:2021xsd}
A.~Mollabashi, N.~Shiba, T.~Takayanagi, K.~Tamaoka and Z.~Wei,
``Aspects of pseudoentropy in field theories,''
Phys. Rev. Res. \textbf{3}, no.3, 033254 (2021)
doi:10.1103/PhysRevResearch.3.033254
[arXiv:2106.03118 [hep-th]].

\bibitem{Mukherjee:2022jac}
J.~Mukherjee,
``Pseudo Entropy in U(1) gauge theory,''
JHEP \textbf{10}, 016 (2022)
doi:10.1007/JHEP10(2022)016
[arXiv:2205.08179 [hep-th]].

\bibitem{Liu:2022ugc}
B.~Liu, H.~Chen and B.~Lian,
``Entanglement Entropy in Timelike Slices: a Free Fermion Study,''
[arXiv:2210.03134 [cond-mat.stat-mech]].

\bibitem{Li:2022tsv}
Z.~Li, Z.~Q.~Xiao and R.~Q.~Yang,
``On holographic time-like entanglement entropy,''
JHEP \textbf{04}, 004 (2023)
doi:10.1007/JHEP04(2023)004
[arXiv:2211.14883 [hep-th]].

\bibitem{He:2023eap}
S.~He, J.~Yang, Y.~X.~Zhang and Z.~X.~Zhao,
``Pseudo-entropy for descendant operators in two-dimensional conformal field theories,''
[arXiv:2301.04891 [hep-th]].

\bibitem{Chen:2023prz}
H.~Y.~Chen, Y.~Hikida, Y.~Taki and T.~Uetoko,
``Complex saddles of three-dimensional de Sitter gravity via holography,''
Phys. Rev. D \textbf{107}, no.10, L101902 (2023)
[arXiv:2302.09219 [hep-th]].

\bibitem{Doi:2023zaf}
K.~Doi, J.~Harper, A.~Mollabashi, T.~Takayanagi and Y.~Taki,
``Timelike entanglement entropy,''
JHEP \textbf{05}, 052 (2023)
doi:10.1007/JHEP05(2023)052
[arXiv:2302.11695 [hep-th]].

\bibitem{Jiang:2023ffu}
X.~Jiang, P.~Wang, H.~Wu and H.~Yang,
``Timelike entanglement entropy and TT\textasciimacron{} deformation,''
Phys. Rev. D \textbf{108}, no.4, 046004 (2023)
doi:10.1103/PhysRevD.108.046004
[arXiv:2302.13872 [hep-th]].

\bibitem{Chen:2023gnh}
Z.~Chen,
``Complex-valued Holographic Pseudo Entropy via Real-time AdS/CFT Correspondence,''
[arXiv:2302.14303 [hep-th]].

\bibitem{Narayan:2023ebn}
K.~Narayan and H.~K.~Saini,
``Notes on time entanglement and pseudo-entropy,''
[arXiv:2303.01307 [hep-th]].

\bibitem{Jiang:2023loq}
X.~Jiang, P.~Wang, H.~Wu and H.~Yang,
``Timelike entanglement entropy in dS$_{3}$/CFT$_{2}$,''
JHEP \textbf{08}, 216 (2023)
doi:10.1007/JHEP08(2023)216
[arXiv:2304.10376 [hep-th]].

\bibitem{Chu:2023zah}
C.~S.~Chu and H.~Parihar,
``Time-like entanglement entropy in AdS/BCFT,''
JHEP \textbf{06}, 173 (2023)
doi:10.1007/JHEP06(2023)173
[arXiv:2304.10907 [hep-th]].

\bibitem{He:2023wko}
S.~He, J.~Yang, Y.~X.~Zhang and Z.~X.~Zhao,
``Pseudo entropy of primary operators in $ T\overline{T}/J\overline{T} $-deformed CFTs,''
JHEP \textbf{09}, 025 (2023)
doi:10.1007/JHEP09(2023)025
[arXiv:2305.10984 [hep-th]].

\bibitem{Chen:2023sry}
H.~Y.~Chen, Y.~Hikida, Y.~Taki and T.~Uetoko,
``Complex saddles of Chern-Simons gravity and dS3/CFT2 correspondence,''
Phys. Rev. D \textbf{108}, no.6, 066005 (2023)
[arXiv:2306.03330 [hep-th]].

\bibitem{Chen:2023eic}
D.~Chen, X.~Jiang and H.~Yang,
``Holographic $T\bar{T}$ deformed entanglement entropy in dS$_3$/CFT$_2$,''
[arXiv:2307.04673 [hep-th]].

\bibitem{Parzygnat:2023avh}
A.~J.~Parzygnat, T.~Takayanagi, Y.~Taki and Z.~Wei,
``SVD Entanglement Entropy,''
[arXiv:2307.06531 [hep-th]].

\bibitem{He:2023ubi}
P.~Z.~He and H.~Q.~Zhang,
``Timelike Entanglement Entropy from Rindler Method,''
[arXiv:2307.09803 [hep-th]].

\bibitem{Guo:2023aio}
W.~z.~Guo and J.~Zhang,
``Sum rule for pseudo R\'enyi entropy,''
[arXiv:2308.05261 [hep-th]].

\bibitem{Omidi:2023env}
F.~Omidi,
``Pseudo R\'enyi Entanglement Entropies For an Excited State and Its Time Evolution in a 2D CFT,''
[arXiv:2309.04112 [hep-th]].

\bibitem{Lewkowycz:2013nqa}
A.~Lewkowycz and J.~Maldacena,
``Generalized gravitational entropy,''
JHEP \textbf{08}, 090 (2013)
doi:10.1007/JHEP08(2013)090
[arXiv:1304.4926 [hep-th]].

\bibitem{Dong:2016hjy}
X.~Dong, A.~Lewkowycz and M.~Rangamani,
``Deriving covariant holographic entanglement,''
JHEP \textbf{11}, 028 (2016)
doi:10.1007/JHEP11(2016)028
[arXiv:1607.07506 [hep-th]].

\bibitem{Dong:2016fnf}
X.~Dong,
``The Gravity Dual of Renyi Entropy,''
Nature Commun. \textbf{7}, 12472 (2016)
doi:10.1038/ncomms12472
[arXiv:1601.06788 [hep-th]].

\bibitem{Dong:2013qoa}
X.~Dong,
``Holographic Entanglement Entropy for General Higher Derivative Gravity,''
JHEP \textbf{01}, 044 (2014)
doi:10.1007/JHEP01(2014)044
[arXiv:1310.5713 [hep-th]].

\bibitem{Casini:2011kv} 
  H.~Casini, M.~Huerta and R.~C.~Myers,
  ``Towards a derivation of holographic entanglement entropy,''
  JHEP {\bf 1105}, 036 (2011)
  [arXiv:1102.0440 [hep-th]].

\bibitem{Callebaut:2023fnf}
N.~Callebaut,
``Entanglement in conformal field theory and holography,''
[arXiv:2303.16827 [hep-th]].

\bibitem{Chen:2020tes}
Y.~Chen, V.~Gorbenko and J.~Maldacena,
``Bra-ket wormholes in gravitationally prepared states,''
JHEP \textbf{02}, 009 (2021)
doi:10.1007/JHEP02(2021)009
[arXiv:2007.16091 [hep-th]].

\bibitem{Goswami:2021ksw}
K.~Goswami, K.~Narayan and H.~K.~Saini,
``Cosmologies, singularities and quantum extremal surfaces,''
JHEP \textbf{03}, 201 (2022)
doi:10.1007/JHEP03(2022)201
[arXiv:2111.14906 [hep-th]].

\bibitem{Narayan:2016xwq} 
  K.~Narayan,
  ``On $dS_4$ extremal surfaces and entanglement entropy in some ghost CFTs,''
  Phys.\ Rev.\ D {\bf 94}, no. 4, 046001 (2016)
  [arXiv:1602.06505 [hep-th]].

\bibitem{Jatkar:2017jwz} 
  D.~P.~Jatkar and K.~Narayan,
  ``Ghost-spin chains, entanglement and $bc$-ghost CFTs,''
  Phys.\ Rev.\ D {\bf 96}, no. 10, 106015 (2017)
  [arXiv:1706.06828 [hep-th]].

\bibitem{Harlow:2011ke} 
  D.~Harlow and D.~Stanford,
  ``Operator Dictionaries and Wave Functions in AdS/CFT and dS/CFT,''
  arXiv:1104.2621 [hep-th].

\bibitem{Maldacena:2019cbz}
J.~Maldacena, G.~J.~Turiaci and Z.~Yang,
``Two dimensional Nearly de Sitter gravity,''
JHEP \textbf{01}, 139 (2021)
doi:10.1007/JHEP01(2021)139
[arXiv:1904.01911 [hep-th]].

\bibitem{Czech:2012bh} 
  B.~Czech, J.~L.~Karczmarek, F.~Nogueira and M.~Van Raamsdonk,
  ``The Gravity Dual of a Density Matrix,''
  Class.\ Quant.\ Grav.\  {\bf 29}, 155009 (2012)
  doi:10.1088/0264-9381/29/15/155009
  [arXiv:1204.1330 [hep-th]].

\bibitem{Wall:2012uf} 
  A.~C.~Wall,
  ``Maximin Surfaces, and the Strong Subadditivity of the Covariant Holographic Entanglement Entropy,''
  Class.\ Quant.\ Grav.\  {\bf 31}, no. 22, 225007 (2014)
  [arXiv:1211.3494 [hep-th]].

\bibitem{Headrick:2014cta} 
  M.~Headrick, V.~E.~Hubeny, A.~Lawrence and M.~Rangamani,
  ``Causality \& holographic entanglement entropy,''
  JHEP {\bf 1412}, 162 (2014)
  doi:10.1007/JHEP12(2014)162
  [arXiv:1408.6300 [hep-th]].

\bibitem{Harlow:2018fse} 
  D.~Harlow,
  ``TASI Lectures on the Emergence of Bulk Physics in AdS/CFT,''
  PoS TASI {\bf 2017}, 002 (2018)
  doi:10.22323/1.305.0002
  [arXiv:1802.01040 [hep-th]].

\bibitem{Headrick:2019eth} 
  M.~Headrick,
  ``Lectures on entanglement entropy in field theory and holography,''
  arXiv:1907.08126 [hep-th].

\bibitem{Almheiri:2014lwa} 
  A.~Almheiri, X.~Dong and D.~Harlow,
  ``Bulk Locality and Quantum Error Correction in AdS/CFT,''
  JHEP {\bf 1504}, 163 (2015)
  doi:10.1007/JHEP04(2015)163
  [arXiv:1411.7041 [hep-th]].

\bibitem{Jafferis:2015del} 
  D.~L.~Jafferis, A.~Lewkowycz, J.~Maldacena and S.~J.~Suh,
  ``Relative entropy equals bulk relative entropy,''
  JHEP {\bf 1606}, 004 (2016)
  doi:10.1007/JHEP06(2016)004
  [arXiv:1512.06431 [hep-th]].

\bibitem{Dong:2016eik} 
  X.~Dong, D.~Harlow and A.~C.~Wall,
  ``Reconstruction of Bulk Operators within the Entanglement Wedge in Gauge-Gravity Duality,''
  Phys.\ Rev.\ Lett.\  {\bf 117}, no. 2, 021601 (2016)
  [arXiv:1601.05416 [hep-th]].

\bibitem{Hayden:2011ag}
P.~Hayden, M.~Headrick and A.~Maloney,
``Holographic Mutual Information is Monogamous,''
Phys. Rev. D \textbf{87}, no.4, 046003 (2013)
doi:10.1103/PhysRevD.87.046003
[arXiv:1107.2940 [hep-th]].

\bibitem{Chandrasekaran:2022cip}
V.~Chandrasekaran, R.~Longo, G.~Penington and E.~Witten,
``An algebra of observables for de Sitter space,''
JHEP \textbf{02}, 082 (2023)
doi:10.1007/JHEP02(2023)082
[arXiv:2206.10780 [hep-th]].

\bibitem{Hartle:1983ai}
J.~B.~Hartle and S.~W.~Hawking,
``Wave Function of the Universe,''
Phys. Rev. D \textbf{28}, 2960-2975 (1983)
doi:10.1103/PhysRevD.28.2960

\bibitem{Bousso:1995cc}
R.~Bousso and S.~W.~Hawking,
``The Probability for primordial black holes,''
Phys. Rev. D \textbf{52}, 5659-5664 (1995)
doi:10.1103/PhysRevD.52.5659
[arXiv:gr-qc/9506047 [gr-qc]].

\bibitem{Bousso:1996au}
R.~Bousso and S.~W.~Hawking,
``Pair creation of black holes during inflation,''
Phys. Rev. D \textbf{54}, 6312-6322 (1996)
doi:10.1103/PhysRevD.54.6312
[arXiv:gr-qc/9606052 [gr-qc]].

\bibitem{Maldacena:2011mk}
J.~Maldacena,
``Einstein Gravity from Conformal Gravity,''
[arXiv:1105.5632 [hep-th]].

\bibitem{Chakraborty:2023yed}
T.~Chakraborty, J.~Chakravarty, V.~Godet, P.~Paul and S.~Raju,
``The Hilbert space of de Sitter quantum gravity,''
[arXiv:2303.16315 [hep-th]].

\bibitem{Holzhey:1994we} 
  C.~Holzhey, F.~Larsen and F.~Wilczek,
  ``Geometric and renormalized entropy in conformal field theory,''
  Nucl.\ Phys.\ B {\bf 424}, 443 (1994)
  [hep-th/9403108].

\bibitem{Calabrese:2004eu} 
  P.~Calabrese and J.~L.~Cardy,
  ``Entanglement entropy and quantum field theory,''
  J.\ Stat.\ Mech.\  {\bf 0406}, P06002 (2004)
  [hep-th/0405152].

\bibitem{Calabrese:2009qy} 
  P.~Calabrese and J.~Cardy,
  ``Entanglement entropy and conformal field theory,''
  J.\ Phys.\ A {\bf 42}, 504005 (2009)
  doi:10.1088/1751-8113/42/50/504005
  [arXiv:0905.4013 [cond-mat.stat-mech]].

\bibitem{VanRaamsdonk:2010pw}
M.~Van Raamsdonk,
``Building up spacetime with quantum entanglement,''
Gen. Rel. Grav. \textbf{42}, 2323-2329 (2010)
doi:10.1142/S0218271810018529
[arXiv:1005.3035 [hep-th]].

\bibitem{VanRaamsdonk:2009ar}
M.~Van Raamsdonk,
``Comments on quantum gravity and entanglement,''
[arXiv:0907.2939 [hep-th]].

\bibitem{Fernandes:2019ige}
K.~Fernandes, K.~S.~Kolekar, K.~Narayan and S.~Roy,
``Schwarzschild de Sitter and extremal surfaces,''
Eur. Phys. J. C \textbf{80}, no.9, 866 (2020)
doi:10.1140/epjc/s10052-020-08437-2
[arXiv:1910.11788 [hep-th]].

\bibitem{Miyaji:2021lcq}
M.~Miyaji,
``Island for gravitationally prepared state and pseudo entanglement wedge,''
JHEP \textbf{12}, 013 (2021)
doi:10.1007/JHEP12(2021)013
[arXiv:2109.03830 [hep-th]].

}
\end{thebibliography}
\end{document}